\documentclass{aa}  
\usepackage{graphicx}
\usepackage{ulem}

\usepackage{txfonts}
\begin{document}

\title{Magnetic reconnection as an erosion mechanism for magnetic switchbacks}

\author{G.H.H. Suen\inst{1} \fnmsep\thanks{\email{ho.suen.20@ucl.ac.uk}} \and C.J. Owen\inst{1} \and D. Verscharen\inst{1} \and T.S. Horbury\inst{2} \and P. Louarn\inst{3} \and R. De Marco\inst{4}}

\institute{Mullard Space Science Laboratory, University College London, Holmbury St. Mary, Dorking, Surrey RH5 6NT, UK
            \and
           Imperial College London, South Kensington Campus, London SW7 2AZ, UK
            \and
           Institut de Recherche en Astrophysique et Plan\'etologie, 9 avenue du Colonel Roche, 31028 Toulouse Cedex 4, France
            \and
          INAF -- Istituto di Astrofisica e Planetologia Spaziali, Via Fosso del Cavaliere 100, 00133, Roma, Italy
            }

\date{Received ; accepted}

\abstract
{Magnetic switchbacks are localised polarity reversals in the radial component of the heliospheric magnetic field. Observations from \textit{Parker Solar Probe} (PSP) have shown that they are a prevalent feature of the near-Sun solar wind. However, observations of switchbacks at 1 au and beyond are less frequent, suggesting that these structures evolve and potentially erode through yet-to-be identified mechanisms as they propagate away from the Sun.}
{We search for magnetic switchbacks undergoing magnetic reconnection, characterise them, and then evaluate the viability of reconnection as a possible channel for their erosion.}
{We analyse magnetic field and plasma data from the Magnetometer and Solar Wind Analyser instruments aboard \textit{Solar Orbiter} between 10 August and 30 August 2021. During this period, the spacecraft was 0.6 -– 0.7 au from the Sun. Using hodographs and Wal\'en analysis methods, we test for rotational discontinuities (RDs) in the magnetic field and reconnection-associated outflows at the boundaries of the identified switchback structures.}
{We identify three instances of reconnection occurring at the trailing edge of magnetic switchbacks, with properties consistent with existing models describing reconnection in the solar wind. Based on these observations, we propose a scenario through which reconnection can erode a switchback and we estimate the timescales over which this occurs. For our events, the erosion timescales are much shorter than the expansion timescale and thus, the complete erosion of all three observed switchbacks would occur well before they reach 1 au. Furthermore, we find that the spatial scale of these switchbacks would be considerably larger than is typically observed in the inner heliosphere if the onset of reconnection occurs close to the Sun. Hence, our results suggest that the onset of reconnection must occur during transport in the solar wind in our cases. These results suggest that reconnection can contribute to the erosion of switchbacks and may explain the relative rarity of switchback observations at 1 au.}
{}


\maketitle


\section{Introduction}
Magnetic reconnection is a fundamental energy conversion process occurring in many laboratory and astrophysical plasmas. Reconnection converts magnetic energy into kinetic and thermal energy through a change in magnetic field topology across current sheets \citep{Pontin2011, Gosling2012, Cassak2016, Hesse2020}. In the context of heliospheric physics, reconnection is a key candidate process to explain coronal heating \citep{Parker1983, Parker1988} and solar wind acceleration \citep{Zank2014, Khabarova2015, Adhikari2019}.

Statistical studies show that up to 20\% of the magnetic energy is converted into particle heating, while the remainder is converted into particle acceleration, creating a pair of oppositely directed bulk outflow jets that stream into the background plasma \citep{Enzl2014, Mistry2017}. The proportion of magnetic energy converted into particle heating and acceleration depends on the magnetic shear angle \citep{Drake2009}. In the rest frame of the reconnection current sheet (RCS), the bulk velocity of the outflow jets is of the order of the local Alfv\'en speed, although reconnection events with sub-Alfv\'enic outflows are not unusual \citep{Haggerty2018, Phan2020}.

The reconnection model described by \citet{Gosling2005a} is frequently used to interpret the spatial structure of reconnection outflows in the solar wind. The RCS bifurcates, forming a pair of standing Alfv\'enic rotational discontinuities (RDs) in the magnetic field at the edges of the outflow region. According to the Rankine-Hugoniot conditions, the RDs must be Alfv\'enic in nature \citep{Hudson1970}. The outflow region has weaker magnetic field strength and increased plasma temperature and density compared to the background plasma. Interplanetary coronal mass ejections (ICMEs) \citep{McComas1994, Gosling2005a}, the heliospheric current sheet (HCS) \citep{Gosling2005b, Gosling2006a, Phan2021}, and the regular solar wind \citep{Gosling2007, Phan2020} are locations where reconnection occurs.

Magnetic switchbacks are localised polarity reversals in the radial component of the heliospheric magnetic field (HMF) \citep{Bale2019, DudokDeWit2020, Krasnoselskikh2020}. They are often associated with increases in the radial component of the bulk proton velocity by a significant fraction of the local Alfv\'en speed \citep{Matteini2014, Horbury2018, Kasper2019, Horbury2020a}. Measurements of suprathermal electrons help us determine the magnetic connectivity during switchback encounters. These electrons stream away from the Sun along the open HMF, forming a field-aligned beam known as the strahl \citep{Owens2013b}. The presence and pitch angle of this beam can be an indicator of the connectivity of open field lines to the Sun \citep{Feldman1975, Rosenbauer1977}: assuming the direction of the strahl velocity indicates the anti-sunward direction along the encountered magnetic field line, the strahl pitch angle in the reversed section of a folded field configuration, such as switchbacks, is the same as in the surrounding HMF \citep{Kasper2019}.

Switchbacks have previously been observed by \textit{Helios} \citep{Horbury2018}, \textit{Ulysses} \citep{Balogh1999}, and \textit{ACE} \citep{Owens2013a} at heliocentric distances between 0.3 -- 2.4 au, both near the ecliptic plane and at high heliolatitudes. Recent observations from \textit{Parker Solar Probe} (PSP) show that switchbacks are a prevalent feature of the near-Sun solar wind \citep{Bale2019, Kasper2019}, present for roughly 75\% of the time during PSP Encounter 1 \citep{Horbury2020a}. These structures are convected over the observing spacecraft on timescales ranging from a few minutes to a few hours \citep{DudokDeWit2020}, and have transverse scales comparable to solar granulation and supergranulation \citep{Fargette2021}.

The mechanisms responsible for the formation of switchbacks are still under debate. Recent studies suggest that at least some proportion of the total switchback population originates from the solar corona \citep{DePablos2022, Telloni2022}. Processes linked to interchange reconnection \citep{Fisk2020, Drake2021} and coronal jets \citep{Sterling2020, Neugebauer2021} are invoked to explain the formation of these structures in the corona. Other studies suggest that switchback formation can occur locally in the solar wind: for example, based on PSP observations, \citet{Schwardron2021} propose a mechanism through which switchbacks are generated by velocity shears between fast and slow solar wind streams. Magnetohydrodynamic (MHD) simulations suggest that Alfv\'en wave steepening \citep{Squire2020, Johnston2022, Squire2022} and Kelvin-Helmholtz instabilities \citep{Ruffolo2020, Kieokaew2021} are viable formation mechanisms.

At heliocentric distances of 1\,au and beyond, switchbacks are less frequently seen than in the inner heliosphere, suggesting that these structures evolve and eventually erode as they propagate away from the Sun \citep{Tenerani2020, Tenerani2021}. Magnetic reconnection is one possible mechanism that can enhance erosion of a switchback by removing magnetic flux from the polarity-reversed section of the magnetic field. Observations from \textit{Helios} \citep{Gosling2006b} and PSP \citep{Froment2021} show that reconnection may occur at switchback boundaries.

We present examples of switchback boundary reconnection events observed by \textit{Solar Orbiter} and use them to evaluate the effectiveness of magnetic reconnection as an erosion mechanism for switchbacks. In Sect. 2, we describe our data and analysis methods. In Sect. 3, we show observations of three instances of switchback reconnection. In Sect. 4, we present our interpretation of the switchback and reconnection geometry based on the observations, and estimate the remaining lifetime of the switchbacks. In Sect. 5, we summarise our findings and discuss their implications on the global properties of switchbacks in the solar wind.

\section{Data and methods}
\subsection{Instrumentation}

We use publicly available magnetic field and plasma data from the Magnetometer (MAG, \citealt{Horbury2020b}) and Solar Wind Analyser (SWA, \citealt{Owen2020}) instruments on board \textit{Solar Orbiter}. The MAG instrument consists of a pair of fluxgate magnetometers mounted on the spacecraft boom and has a measurement cadence of 8 vectors/second in normal mode operation during this period. The SWA instrument is comprised of three sensors --- of particular relevance to this study are the SWA-Proton Alpha Sensor (SWA-PAS) and SWA-Electron Analyser System (SWA-EAS). SWA-PAS delivers ground-calculated proton moments (velocity, temperature, and density) once every 4 seconds for periods of normal mode operation. We use electron strahl pitch angle distribution (PAD) data at $>70$ eV from SWA-EAS, when available, at a cadence of one measurement per 10 seconds.

For this case study, we sample a time interval during August 2021 for magnetic reconnection outflows in the solar wind, when the spacecraft was at a heliocentric distance of 0.6 -- 0.7\,au. We exclude outflows with crossing durations less than 20 seconds to ensure that there are at least five proton measurements inside the outflow region. Out of the ten events that satisfy the selection criteria, three are associated with potential magnetic switchbacks.

\subsection{Reference frames}
The MAG and SWA data are initially provided in the RTN coordinate system. This is a spacecraft-centred reference frame in which $\mathbf{\hat{R}}$ is the Sun-spacecraft radial vector, $\mathbf{\hat{T}}$ is the cross product between the Sun's rotation axis and $\mathbf{\hat{R}}$, and $\mathbf{\hat{N}}$ completes the triad. 

We rotate vector quantities into a current sheet-aligned $(lmn)$-frame, defined by the basis vectors $\mathbf{\hat{l}}$, $\mathbf{\hat{m}}$, and $\mathbf{\hat{n}}$. This coordinate system is derived using the hybrid minimum variance analysis (MVAB) method \citep{Gosling2013}. We calculate the current sheet normal direction $\mathbf{\hat{n}}$ as:
\begin{equation}
    \hat{\mathbf{n}} = \frac{\mathbf{B_1}\times\mathbf{B_2}}{|\mathbf{B_1}\times\mathbf{B_2}|},
    \label{eq1}      
\end{equation}
where $\mathbf{B_{1}}$ and $\mathbf{B_{2}}$ are the instantaneous magnetic field vectors on either side of the current sheet. $\mathbf{\hat{m}}$ is given by $\hat{\mathbf{m}} = \hat{\mathbf{l^\prime}} \times \hat{\mathbf{n}}$, where $\mathbf{\hat{l^\prime}}$ is the maximum variance direction unit vector derived from the standard MVAB method developed by \citet{Sonnerup1967}. $\mathbf{\hat{l}}$ completes the triad as $\hat{\mathbf{l}} = \hat{\mathbf{m}} \times \hat{\mathbf{n}}$. Since the MVAB method requires the solution of an eigenvalue problem, the reliability of $\mathbf{\hat{l^\prime}}$ (and hence, $\hat{\mathbf{m}}$) depends on non-degeneracy in the eigenvalues corresponding to the maximum ($\lambda_1$) and intermediate ($\lambda_2$) variance direction eigenvectors. Typically, this non-degeneracy condition is satisfied if $\lambda_1/\lambda_2\geq10$ \citep{Sonnerup1998}. Assuming the current sheet is planar and $B_n\ll B_m$, we expect  $\mathbf{\hat{l}}$, $\mathbf{\hat{m}}$, and $\mathbf{\hat{n}}$ to broadly correspond to the exhaust outflow direction, the x-line direction and current sheet normal direction, respectively. 

\begin{figure*}[h!]
   \centering
   \includegraphics[width=\textwidth]{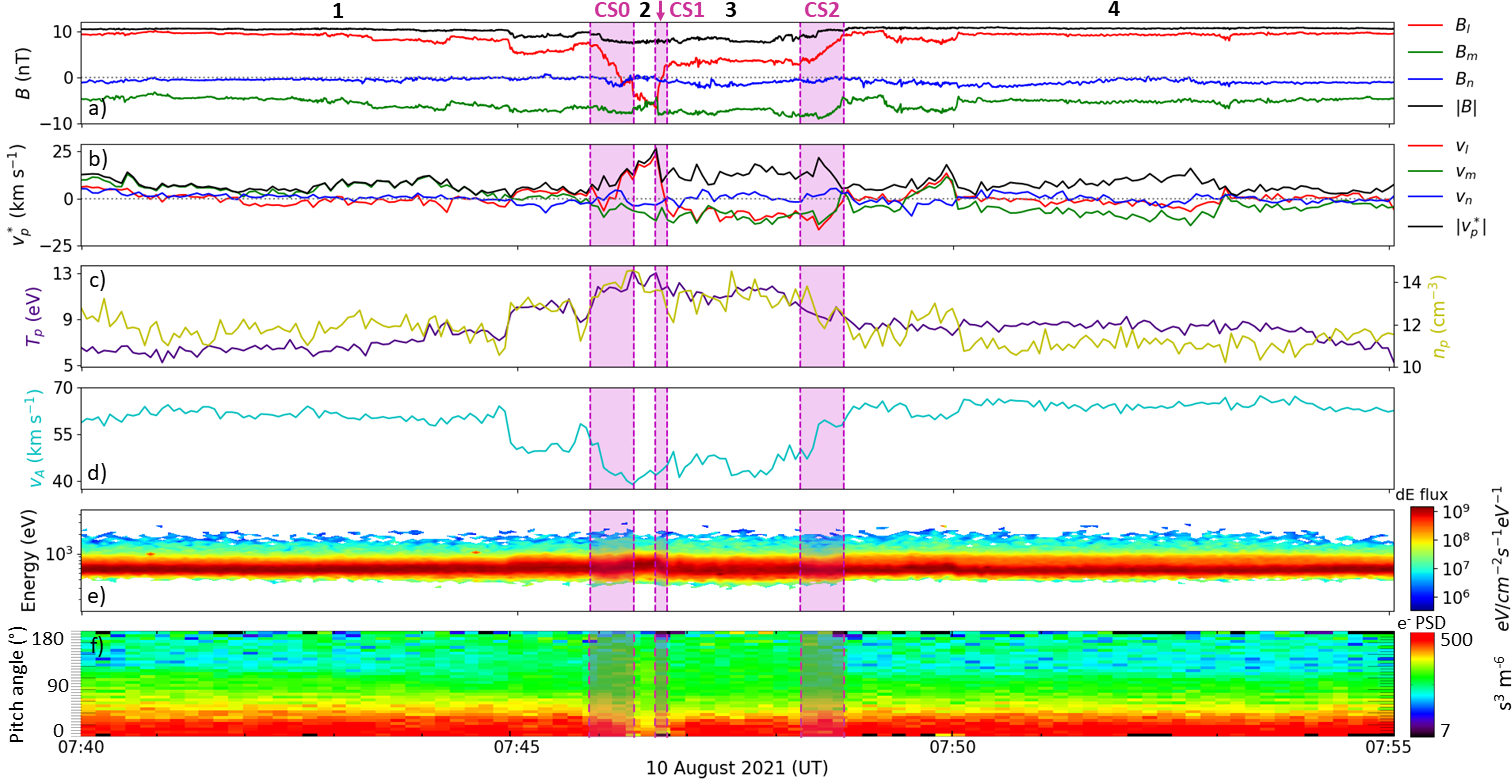}
   \caption{Combined magnetic field, proton, and electron strahl PAD time series data for Event 1 in the hybrid MVAB $lmn$-frame. a) Magnetic field vector with the magnetic field strength in black. b) Proton bulk velocity with the proton bulk speed in black. The average proton bulk velocity $\langle\mathbf{v_p}\rangle$ over this interval has been removed. In both panels, the $l$-component is in red, the $m$-component is in green, and the $n$-component is in blue. c) Proton temperature (left scale, purple) and number density (right scale, gold). d) Alfv\'en speed. e) 1D proton energy spectrogram. f) Electron strahl PAD for energies $>70$ eV. The dashed lines mark the region boundaries identified in the text and numbered at the top of the figure.}
    \label{fig1}
\end{figure*}

The deHoffman-Teller (HT) frame is defined as the reference frame in which the convection electric field vanishes \citep{deHoffmann1950}. We find the RTN HT frame velocity $\mathbf{v_{HT}}$ by solving the linear matrix equation \citep{Paschmann2008} over a given measurement interval
\begin{multline}    
    \begin{pmatrix}
        \langle{B_T^2 + B_N^2}\rangle & \langle{-B_RB_T}\rangle & \langle{-B_RB_N}\rangle \\
        \langle{-B_RB_T}\rangle & \langle{B_R^2 + B_N^2}\rangle & \langle{-B_TB_N}\rangle \\
        \langle{-B_RB_N}\rangle & \langle{-B_TB_N}\rangle & \langle{B_R^2 + B_T^2}\rangle
    \end{pmatrix}
    \begin{pmatrix}
        v_{HT,R} \\
        v_{HT,T} \\
        v_{HT,N}
    \end{pmatrix} = \\
     \begin{pmatrix}
        \langle{E_TB_N - E_NB_T}\rangle \\
        \langle{E_NB_R - E_RB_N}\rangle \\
        \langle{E_RB_T - E_TB_R}\rangle \\
    \end{pmatrix},
    \label{eq2}
\end{multline}
where $\mathbf{B}$ is the magnetic field vector, $\mathbf{E}$ is the convection electric field, and the angled brackets $\langle\rangle$ denote averages over the measurements used in this calculation. In the ideal MHD limit, we assume that $\mathbf{E}=-\mathbf{v_p}\times\mathbf{B}$ where $\mathbf{v_p}$ is the RTN frame bulk plasma velocity.

\subsection{Testing for rotational discontinuities}
Magnetic hodographs are used to illustrate the spatial and temporal evolution of $\mathbf{B}$ in 3D; they are plotted in pairs for the $lm$ and $ln$-planes of the $lmn$-frame \citep{Sonnerup1998}. If an RD is present across the current sheet, we expect to see the temporal variation of $\mathbf{B}$ trace a semi-circular arc in the $lm$-plane hodograph and a vertical line at $B_n\neq0$ in the $ln$-plane hodograph.

We use hodographs in conjunction with the Wal\'en relation to test for Alfv\'enic RDs across current sheets \citep{Khrabrov1998}:
\begin{equation}
    \mathbf{v_p^\prime} = \pm\mathbf{v_A} = \pm\frac{\mathbf{B}}{\sqrt{\mu_0\rho}},
    \label{eq3}
\end{equation}  
where $\mathbf{v_p^\prime} = \mathbf{v_p} - \mathbf{v_{HT}}$ is the HT frame bulk plasma velocity, $\mu_0$ is the permittivity of free space, and $\rho$ is the plasma mass density. The sign in Eq. \ref{eq3} indicates whether the Alfv\'enic fluctuations in $\mathbf{v_p^\prime}$ and $\mathbf{B}$ are correlated (positive) or anti-correlated (negative). We test the strength of the Wal\'en relation using component-by-component scatter plots of $\mathbf{v_p^\prime}$ against $\mathbf{v_A}$, using the least-squares linear regression method to determine the line of best fit. From Eq. \ref{eq3}, a line of best fit slope of $\pm1$ in the Wal\'en plot is an indicator of an ideal Alfv\'enic RD, although previous works suggest that slopes with magnitudes between 0.5 -- 1 are sufficient to demonstrate the existence of an RD across a current sheet \citep{Paschmann2005, Dong2017}.

\section{Results}
\subsection{Event 1 -- 10 August 2021 07:45:50 - 07:48:45 UT}
Figure \ref{fig1} provides a general overview of the magnetic field and solar wind conditions observed between 07:40:00 and 07:55:00 UT on 10 August 2021, recorded at a heliocentric distance of 0.72\,au. Panel a) shows the magnetic field $\mathbf{B}$ and b) shows the proton bulk velocity $\mathbf{v_p^*}$ in the $lmn$-frame. In both panels, the $l$-component is in red, the $m$-component is in green, and the $n$-component is in blue. Panel c) shows the proton temperature $T_p$ in purple and proton number density $n_p$ in gold, d) shows the Alfv\'en speed $v_A$, e) shows the 1D proton energy spectrogram, and f) shows the electron strahl PAD for energies >70\,eV. We remove the average proton bulk velocity $\mathbf{\langle v_p\rangle}$ across this interval from the data, such that $\mathbf{v_p^*=v_p-\langle v_p\rangle}$.

\begin{table}[h!] 
    \caption{Event 1 $lmn$-frame basis vectors for CS0 and CS1 + CS2 expressed in RTN coordinates.}
    \centering
    $\begin{array}{p{0.40\linewidth}l}
        \hline
        \noalign{\smallskip}
        Current sheet &  lmn$-frame basis vectors (R, T, N)$ \\
        \noalign{\smallskip}
        \hline
        \noalign{\smallskip}
         & \mathbf{\hat{l}} = (0.660, -0.216, 0.719)\\
        CS0 & \mathbf{\hat{m}} = (0.016, -0.954, -0.300)\\
         & \mathbf{\hat{n}} = (0.751, 0.210, -0.626)\\
         \noalign{\smallskip}
         \hline
         \noalign{\smallskip}
         & \mathbf{\hat{l}} = (0.677, -0.343, 0.651)\\
        CS1 + CS2 & \mathbf{\hat{m}} = (0.086, 0.916, 0.393)\\
         & \mathbf{\hat{n}} = (-0.731, -0.210, 0.649)\\         
        \noalign{\smallskip}
        \hline
    \end{array}$
    \label{tab1} 
\end{table}

\begin{figure*}[h!]
   \centering
   \includegraphics[width=\textwidth]{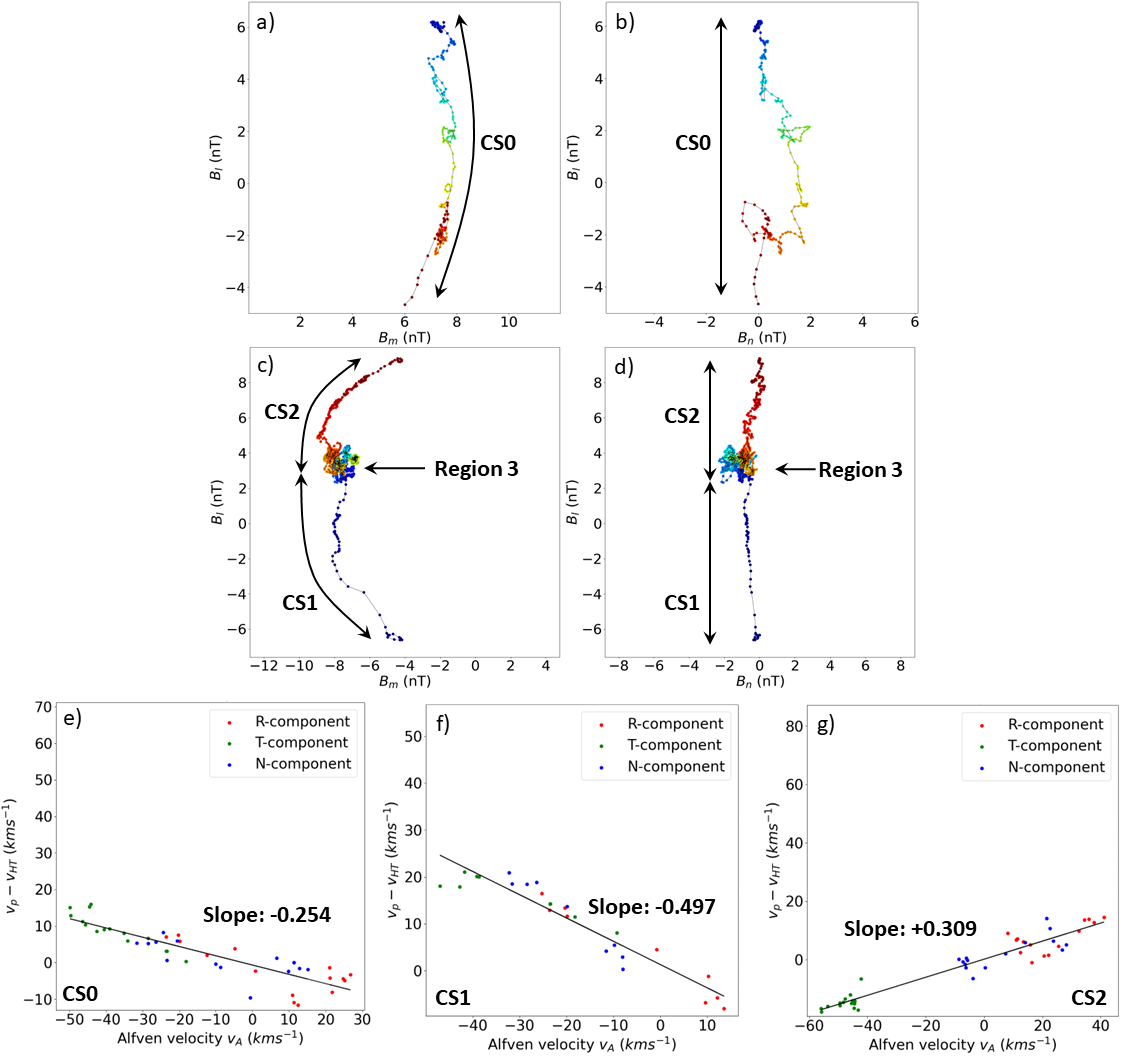}
   \caption{Magnetic hodographs and Wal\'en plots for CS0 (07:45:50 -- 07:46:20 UT), CS1 (07:46:35 -- 07:46:43 UT), and CS2 (07:48:15 -- 07:48:45 UT) in Event 1. Time progression in the hodographs is represented by the colour of the dots, with earlier times in blue and later times in red. The red, green, and blue dots in the Wal\'en plots represent the $R$, $T$, and $N$-components of the Alfv\'en velocity $\mathbf{v_A}$ and the HT frame bulk plasma velocity $\mathbf{v_p-v_{HT}}$. a) $lm$-plane hodograph for CS0. b) $ln$-plane hodograph for CS0. c) $lm$-plane hodograph for CS1 and CS2. d) $ln$-plane hodograph for CS1 and CS2. e) Wal\'en plot for CS0. f) Wal\'en plot for CS1. g) Wal\'en plot for CS2.}
    \label{fig2}
\end{figure*}

In RTN coordinates, $\mathbf{\langle v_p\rangle} = (322.2, -5.6, -5.6)_{RTN}$\,km s$^{-1}$ across this time interval and the predominant HMF polarity is in the anti-sunward $(+R)$ direction. We divide this interval into several regions marked by the vertical dashed lines. Regions 1 (07:40:00 -- 07:45:50 UT) and 4 (07:48:45 -- 07:55:00 UT) correspond to the period of quiet HMF and steady, slow solar wind surrounding this event. The regions shaded in purple are centered around sharp discontinuities in the magnetic field that we identify as current sheets.

We derive the $lmn$-frames for the current sheets at the leading (CS0) and trailing edges (CS1, CS2) of this event using the hybrid MVAB method. As the trailing edge current sheets are bifurcated, we perform the MVAB analysis from the start of CS1 to the end of CS2. Table 1 shows the $lmn$-frame basis vectors for these current sheets. The angular differences between the corresponding basis vector pairs of both frames are small, ranging from $1.7^\circ$ to $8.3^\circ$. Thus, the $lmn$-frames for the leading and trailing edges of this event are roughly aligned. As we are interested in the properties of the reconnection outflow, we visualise its properties in the $lmn$-frame of the trailing edge current sheet in Figs. \ref{fig1}a and \ref{fig1}b. We do the same for the overview plots of the other two events.

\begin{figure*}[h!]
   \centering
   \includegraphics[width=\textwidth]{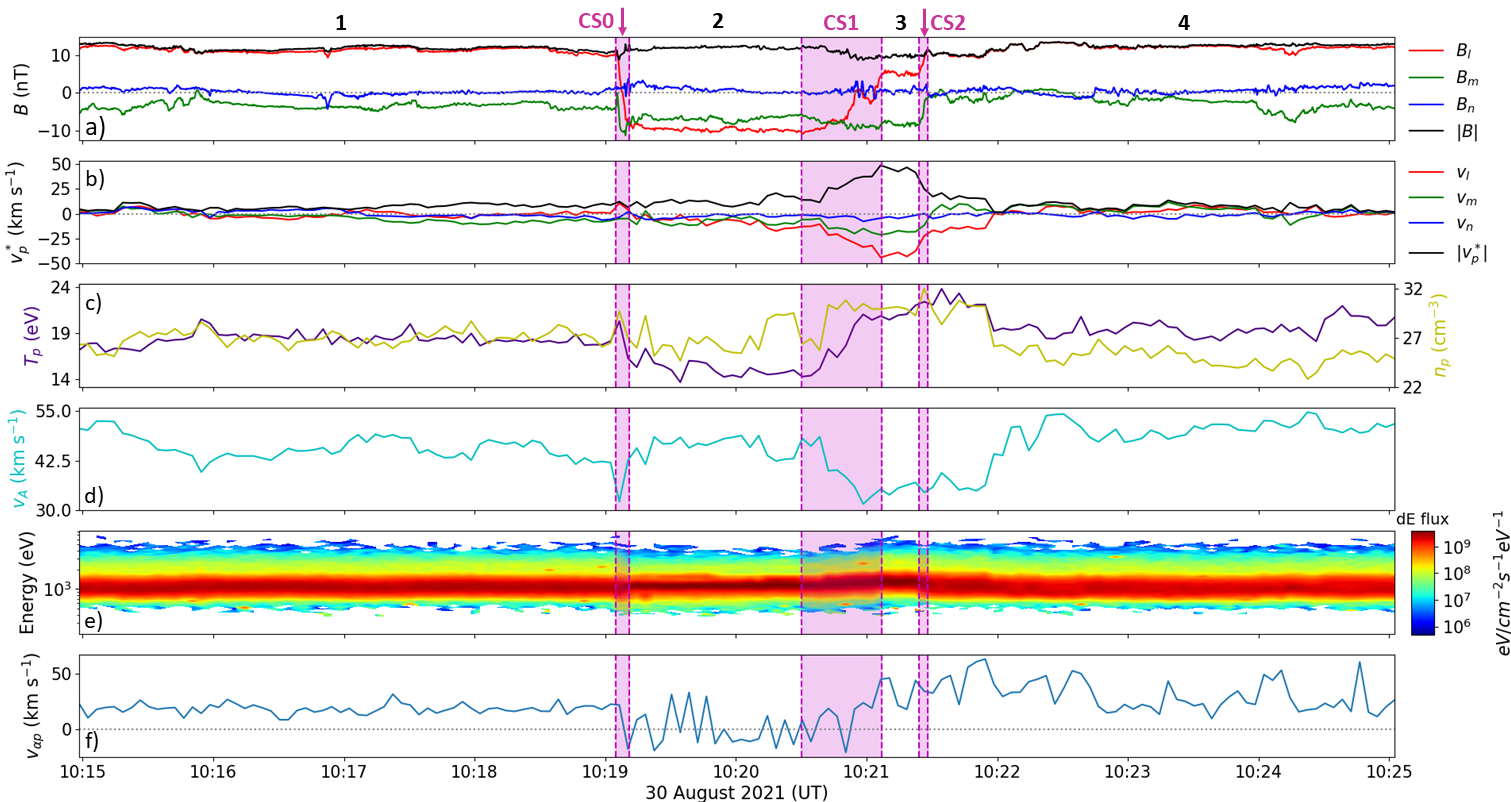}
   \caption{Combined magnetic field and proton time series data for Event 2 in the hybrid MVAB $lmn$-frame. The figure layout is the same as in Figure \ref{fig1} except for the absence of electron strahl PAD data, which are unavailable for this interval. Panel f) instead shows the signed magnitude of the alpha-proton velocity difference vector $v_{\alpha p}$.}
    \label{fig3}
\end{figure*}

Across CS0 (07:45:50 -- 07:46:20 UT), the polarity of the radial component of the HMF, $B_R$, flips from the anti-sunward direction to the sunward direction. In the $lmn$-frame of this event, this corresponds to a reversal in the $B_l$ component of the magnetic field from +7\,nT to -4\,nT. Due to the relatively strong $B_m$ component, the maximum magnetic shear angle across this current sheet is $77.2^{\circ}$. There is a 20\% decrease in the average magnetic field strength $|\mathbf{B}|$, from 10\,nT in Region 1 to 8\,nT in CS0. $v_l$, the $l$-component of $\mathbf{v_p^*}$, increases from 0\,km s$^{-1}$ to +10\,km\,s$^{-1}$, and the average proton bulk speed $|\mathbf{v_p^*}|$ increases from 4\,km s$^{-1}$ to 13\,km s$^{-1}$. Here, we also measure the maximum $T_p$ of 13\,eV and $n_p$ of 14\,cm$^{-3}$.

Region 2 (07:46:20 -- 07:46:35 UT) encompasses the polarity-reversed section of this event. $B_l$ decreases to -6\,nT, $v_l$ increases further to +24\,km\,s$^{-1}$ and $|\mathbf{v_p^*}|$ increases to +27\,km\,s$^{-1}$. This is roughly $68\%$ of the local $v_A$ of 40\,km\,s$^{-1}$. There is minimal change in $|\mathbf{B}|$, $T_p$, and $n_p$ in this region compared to CS0. The electron strahl PAD peaks in the field-aligned direction ($0^{\circ})$ both in the background HMF and in the regions containing polarity-reversed magnetic flux (CS0 and Region 2).

Across CS1 (07:46:35 -- 07:46:43 UT) and CS2 (07:48:15 -- 07:48:45 UT), the HMF polarity reverts back towards the anti-sunward direction observed in Region 1. $B_l$ increases from -6\,nT to +3\,nT across CS1 and then increases again from +3\,nT to +9\,nT across CS2. In Region 3 (07:46:43 -- 07:48:15 UT), $B_l$ remains roughly constant at +4\,nT; this is intermediate between its value in Region 2 and the background HMF in Regions 1 and 4. The total magnetic shear angle across CS1, CS2, and Region 3 is $117^{\circ}$. There is a slight decrease in $|\mathbf{v_p^*}|$ from 25\,km s$^{-1}$ to an average of 15\,km s$^{-1}$. $v_l$ sharply decreases across CS1 and is negative in Region 3, with an average value of -10\.km\,s$^{-1}$. We observe gradual decreases in $T_p$ from 12.5\,eV to 9\,eV and in $n_p$ from 14\,cm$^{-3}$ to 12\,cm$^{-3}$. Moreover, we note a brief strahl dropout across CS1 and the latter part of Region 2, accompanied by a sustained broadening of the strahl PAD in CS1 and Region 3. As both features are also present in the raw electron counts data, they are unlikely to be aliasing effects caused by the rapid rotation of the magnetic field.

Using the methods described in Section 2.3, Fig. \ref{fig2} shows the magnetic hodographs and Wal\'en plots for CS0, CS1, and CS2. Panels a) and b) show the hodographs for CS0 in its associated $lmn$-frame, panels c) and d) show the hodographs for CS1 and CS2 combined in their associated $lmn$-frame, and panels e) -- g) show the Wal\'en plots for CS0, CS1, and CS2. For the Wal\'en plots, we re-sample $\mathbf{B}$ on $\mathbf{v_p}$ as MAG has higher time resolution than PAS. We also include all data points 15 seconds before and after the current sheet crossing in the analysis. This ensures that a representative number of data points are included in the Wal\'en plots, even for short-duration current sheets containing only a single proton measurement inside the current sheet. The choice of 15 seconds is deliberate, to prevent data points from CS1 contaminating the analysis for CS0 and vice-versa. For consistency, we apply the same method and the same timeframe of 15 seconds to all three events.

\begin{figure*}[t!]
   \centering
   \includegraphics[width=0.9\textwidth]{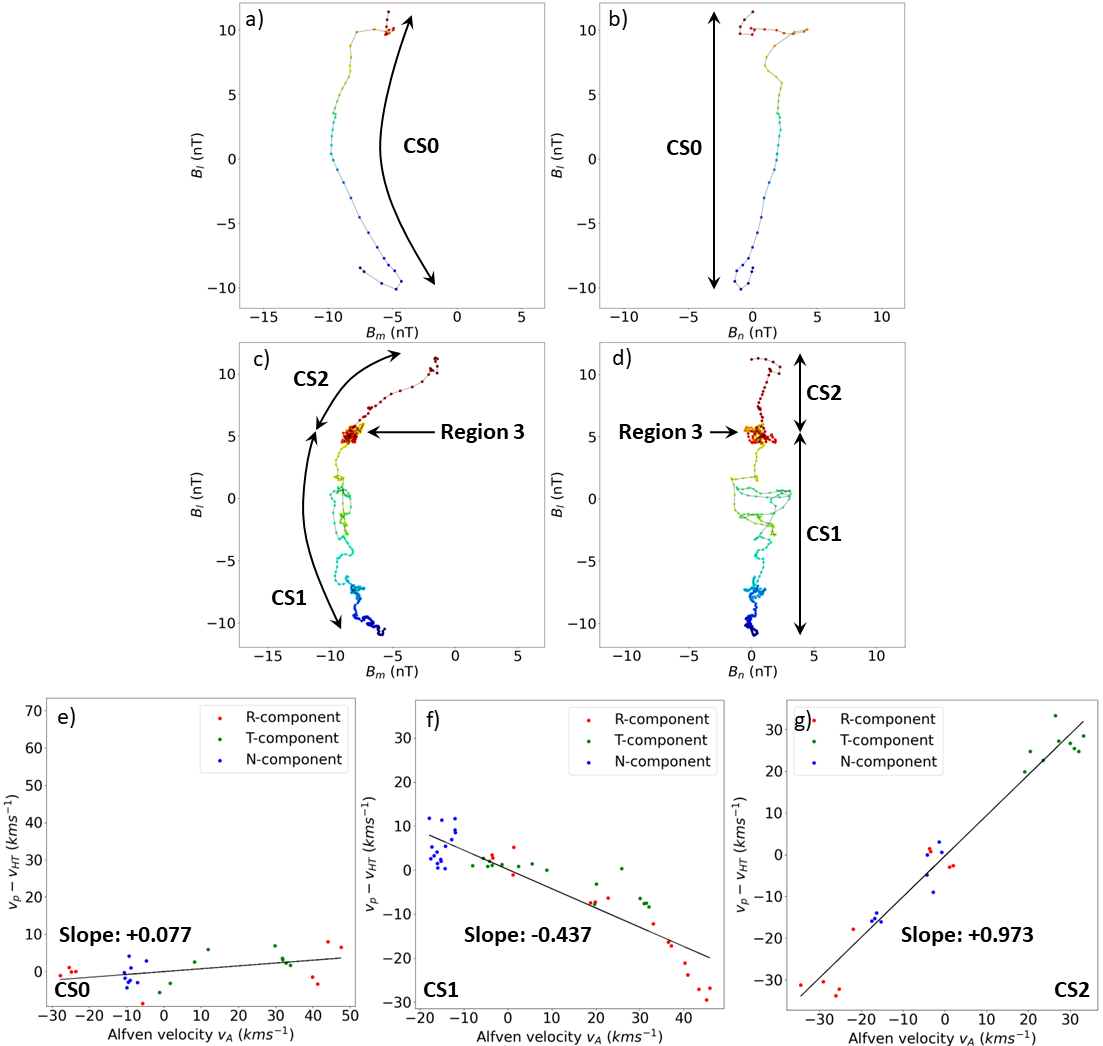}
   \caption{Magnetic hodographs and Wal\'en plots for CS0 (10:19:05 -- 10:19:11 UT), CS1 (10:20:50 -- 10:21:07 UT), and CS2 (10:21:24 -- 10:21:28 UT) in Event 2. The figure layout is the same as in Figure \ref{fig2}.}
    \label{fig4}
\end{figure*}

In the $lm$-plane hodographs (Figs. \ref{fig2}a, \ref{fig2}c), $\mathbf{B}$ across all three current sheets traces an arc consistent with the measured magnetic shear angle. In the $ln$-plane hodograph for CS0 (Fig. \ref{fig2}b), $B_n\simeq0$\,nT at the start and end of the interval, but deflects out to $B_n\simeq+2$\,nT in the middle. For CS1 and CS2 (Fig. \ref{fig2}d), $\mathbf{B}$ has a small $B_n$ component of $-1.0$\,nT and traces a quasi-vertical line in the $ln$-plane. In Figs. \ref{fig2}c and \ref{fig2}d, the rotation of $\mathbf{B}$ is split into two arcs that individually correspond to CS1 and CS2. They are separated by an interval where the orientation of $\mathbf{B}$ does not change significantly, corresponding to Region 3. The magnitudes of the gradient of the line of best fit of the Wal\'en plots for CS0 (-0.254), CS1 (-0.497), and CS2 (+0.309) fall below the range $0.5$ -- $1$ expected for an Alfv\'enic structure.

\subsection{Event 2 - 30 August 2021 10:19:05 - 10:21:28 UT}
Figure \ref{fig3} shows Event 2 observed between 10:15:00 and 10:25:00 UT on 30 August 2021 at a heliocentric distance of 0.61 au. The figure layout is the same as in Fig. \ref{fig1}, except for the absence of electron strahl PAD data. In lieu of the strahl PAD, panel f) instead shows the signed magnitude of the alpha-proton velocity difference vector $v_{\alpha p} = |\mathbf{v_{\alpha}-v_p}|\cdot$sgn$(v_{\alpha,R}-v_{p,R})$, which we use as an alternative method of checking for folded field configurations \citep{Fedorov2021}. We obtain this data using the techniques described in \citet{DeMarco2023}. For this interval, $\mathbf{\langle v_p\rangle} = (438.8, -14.6, -2.3)_{RTN}$\,km\,s$^{-1}$ and the predominant HMF polarity before (Region 1, 10:15:00 -- 10:19:05 UT) and after (Region 4, 10:21:28 -- 10:25:00 UT) this event is in the sunward direction. 

We again identify three regions of strong magnetic gradients and label them CS0, CS1, and CS2. Table \ref{tab2} shows the $lmn$-frame basis vectors for these current sheets. The angular differences between the basis vectors of the $lmn$-frames for CS0 and CS1 + CS2 are $21.6^\circ$ for $\mathbf{\hat{l}}$, $28.4^\circ$ for $\mathbf{\hat{m}}$, and $18.9^\circ$ for $\mathbf{\hat{n}}$.

\begin{table}[h!] 
    \caption{Event 2 $lmn$-frame basis vectors for CS0 and CS1 + CS2 expressed in RTN coordinates.}
    \centering
    $\begin{array}{p{0.40\linewidth}l}
        \hline
        \noalign{\smallskip}
        Current sheet &  lmn$-frame basis vectors (R, T, N)$ \\
        \noalign{\smallskip}
        \hline
        \noalign{\smallskip}
         & \mathbf{\hat{l}} = (0.971, -0.220, -0.093)\\
        CS0 & \mathbf{\hat{m}} = (-0.185, -0.939, 0.287)\\
         & \mathbf{\hat{n}} = (-0.151, -0.261, -0.953)\\
         \noalign{\smallskip}
         \hline
         \noalign{\smallskip}
         & \mathbf{\hat{l}} = (-0.835, 0.550, -0.020)\\
        CS1 + CS2 & \mathbf{\hat{m}} = (-0.452, -0.665, 0.594)\\
         & \mathbf{\hat{n}} = (0.313, 0.505, 0.804)\\         
        \noalign{\smallskip}
        \hline
    \end{array}$
    \label{tab2} 
\end{table}

\begin{figure*}[h!]
   \centering
   \includegraphics[width=\textwidth]{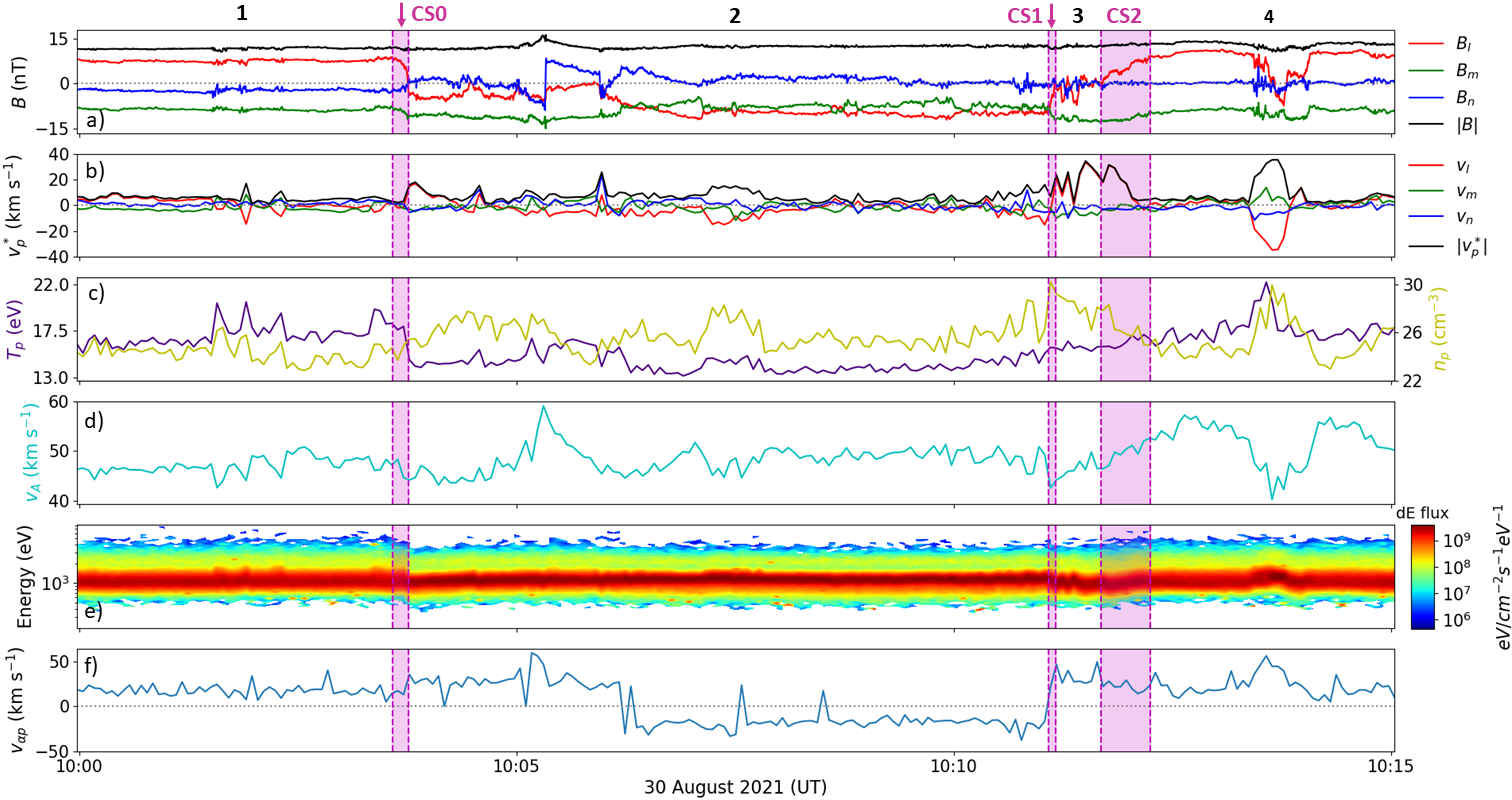}
   \caption{Combined magnetic field and proton time series data for Event 3 in the hybrid MVAB $lmn$-frame. The figure layout is the same as Figure \ref{fig3}.}
    \label{fig5}
\end{figure*}

$B_R$ flips from its sunward-orientation in Region 1 to an anti-sunward orientation in Region 2 (10:19:11 -- 10:20:30 UT) across CS0 (10:19:05 -- 10:19:11 UT). In the $lmn$-frame, this is visible as a reversal in $B_l$ from +10\,nT to -9\,nT; the maximum magnetic shear angle across this current sheet is $113^\circ$. There are no major changes in $|\mathbf{B}|$, $v_l$, and $|\mathbf{v_p^*}|$ in this region from their values in the background HMF in Region 1. $T_p$ decreases from 19\,eV to 14\,eV and there is a brief spike in $n_p$ up to a value of 29\,cm$^{-3}$.

Region 2 corresponds to the polarity-reversed section of this event. $B_l$ and $|\mathbf{B}|$ both remain approximately constant at -10\,nT and 12\,nT, respectively. $v_l$ decreases gradually over Region 2 from 0\,km\,s$^{-1}$ to -14\,km\,s$^{-1}$, while there is a very slight increase in $|\mathbf{v_p^*}|$ from 7\,km\,s$^{-1}$ to 14\,km\,s$^{-1}$. This is around 30\% of the average local $v_A\sim45$\,km\,s$^{-1}$. We measure a roughly constant average $T_p$ of 14\,eV and fluctuations in $n_p$ about an average value of 25\,cm$^{-3}$.

$B_l$ reverses from -10\,nT to +10\,nT in two steps across CS1 (10:20:30 -- 10:21:07 UT) and CS2 (10:21:24 -- 10:21:28 UT), dwelling at +5\,nT in Region 3 (10:21:07 -- 10:21:24 UT). The total magnetic shear across CS1 and CS2 is $134^\circ$ and $|\mathbf{B}|$ decreases from 12.5\,nT to 10\,nT. Across CS1, $v_l$ continues decreasing at a faster rate than in Region 2, reaching a minimum value of -45\,km\,s$^{-1}$ in Region 3. $|\mathbf{v_p^*}|$ peaks at 50\,km\,s$^{-1}$, a value $\sim43\%$ greater than the local $v_A\sim35$\,km\,s$^{-1}$. We observe increases in $T_p$ from 14\,eV to 25\,eV and $n_p$ from 25\,cm$^{-3}$ to 30\,cm$^{-3}$.

Figure \ref{fig4} shows the hodographs and Wal\'en plots for Event 2, the format of this figure is the same as in Fig. \ref{fig2}. The arc traced by $\mathbf{B}$ in the $lmn$-frame for CS0, CS1, and CS2 is consistent with the measured magnetic shear. Across the trailing edge current sheets, the largest rotation in $\mathbf{B}$ occurs over CS1. In the $ln$-plane, $\mathbf{B}$ traces an approximately vertical line and has a $B_n$ component of +0.5\,nT. Around 10:21:00, there are fluctuations in $B_n$ of $\pm2.5$\,nT inside CS1. The Wal\'en plot gradients of +0.077 for CS0 and -0.437 for CS1 are below the range expected for an Alfv\'enic RD. Conversely, the Wal\'en plot gradient of +0.973 for CS2 indicates that the discontinuity in $\mathbf{B}$ across this structure is Alfv\'enic.

\subsection{Event 3 - 30 August 2021 10:03:46 - 10:12:15 UT}
Event 3 (Fig. \ref{fig5}) is observed between 10:00:00 UT to 10:15:00 UT on 30 August when \textit{Solar Orbiter} was at a heliocentric distance of 0.61 au. This event has a duration of around eight minutes, four times greater than that of Events 1 and 2. Event 3 occurs in close temporal proximity to Event 2, thus $\mathbf{\langle v_p\rangle}$ and the background HMF polarity for both events are similar. Table \ref{tab3} shows the $lmn$-frame basis vectors for CS0 and CS1 + CS2. The angular differences between the basis vectors of these $lmn$-frames are $24.6^\circ$ for $\mathbf{\hat{l}}$, $20.2^\circ$ for $\mathbf{\hat{m}}$, and $13.9^\circ$ for $\mathbf{\hat{n}}$.

\begin{table}[h!] 
    \caption{Event 3 $lmn$-frame basis vectors for CS0 and CS1 + CS2 expressed in RTN coordinates.}
    \centering
    $\begin{array}{p{0.40\linewidth}l}
        \hline
        \noalign{\smallskip}
        Current sheet &  lmn$-frame basis vectors (R, T, N)$ \\
        \noalign{\smallskip}
        \hline
        \noalign{\smallskip}
         & \mathbf{\hat{l}} = (-0.970, -0.240, 0.037)\\
        CS0 & \mathbf{\hat{m}} = (-0.235, 0.889, -0.394)\\
         & \mathbf{\hat{n}} = (0.061, -0.391, -0.918)\\
         \noalign{\smallskip}
         \hline
         \noalign{\smallskip}
         & \mathbf{\hat{l}} = (-0.977, 0.177, 0.120)\\
        CS1 + CS2 & \mathbf{\hat{m}} = (-0.115, -0.908, 0.403)\\
         & \mathbf{\hat{n}} = (0.180, 0.308, 0.907)\\         
        \noalign{\smallskip}
        \hline
    \end{array}$
    \label{tab3} 
\end{table}

\begin{figure*}[t!]
   \centering
   \includegraphics[width=0.9\textwidth]{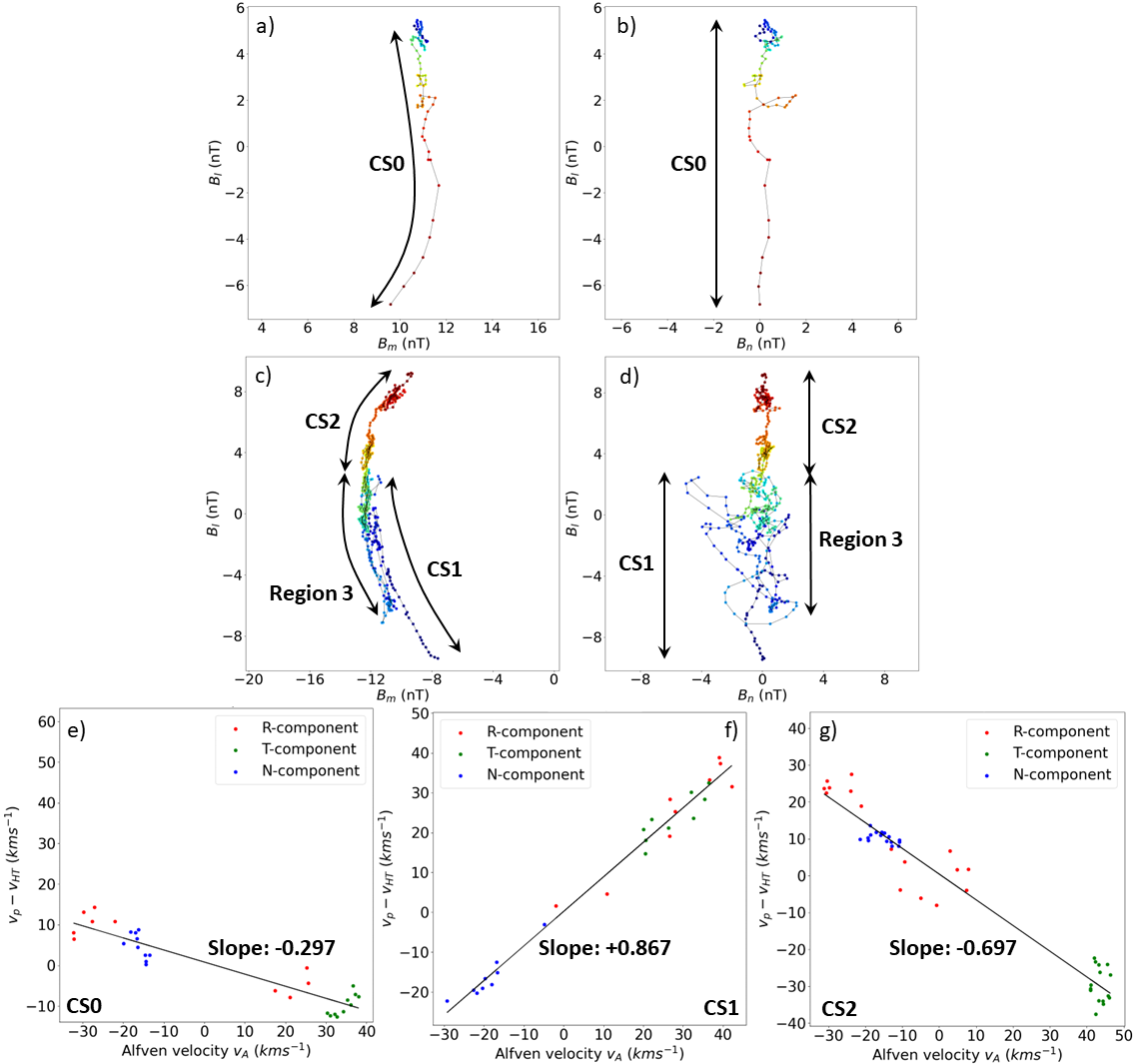}
   \caption{Magnetic hodographs and Wal\'en plots for CS0 (10:03:35 -- 10:03:46 UT), CS1 (10:11:05 -- 10:11:10 UT), and CS2 (10:11:41 -- 10:12:15 UT) in Event 3. The figure layout is the same as in Figure \ref{fig2}.}
    \label{fig6}
\end{figure*}

$B_R$ in Region 2 (10:03:46 -- 10:11:05 UT) is in the anti-sunward direction, opposite to the polarity of the background HMF in Regions 1 (10:00:00 -- 10:03:35 UT) and 4 (10:12:15 -- 10:15:00 UT). This polarity reversal occurs across CS0 (10:03:35 -- 10:03:46 UT), where $B_l$ reverses from +8\,nT to -4\,nT with a magnetic shear angle of $61^\circ$. $|\mathbf{B}|$, $v_l$, and $|\mathbf{v_p^*}|$ do not deviate noticeably from their values in Region 1. $T_p$ decreases from 17\,eV to 14\,eV while there is a slight increase in $n_p$ from 23.5\,cm$^{-3}$ to 25.5\,cm$^{-3}$. 

From 10:03:46 -- 10:05:20, $B_l$ remains roughly constant at -5\,nT and decreases further to -10\,nT from 10:05:55 UT onwards. $|\mathbf{B}|$ is roughly constant at 12\,nT throughout most of Region 2 and is similar in magnitude to $|\mathbf{B}|$ in the background HMF. There is an increase in $|\mathbf{B}|$ between 10:05:20 and 10:05:55 UT that coincides with near-zero $B_l$ and a large deflection in $B_n$, suggesting that this event contains some internal substructure that is not evident in the other two events. There is no significant change in $v_l$ or $|\mathbf{v_p^*}|$ in Region 2. $T_p$ for the full duration of Region 2 is roughly constant at 13\,eV, while $n_p$ increases gradually from 25.5\,cm$^{-3}$ to 30\,cm$^{-3}$.

Across CS1 (10:11:05 -- 10:11:10 UT), $B_l$ increases from -10\,nT to -2\,nT and then reverses from -1\,nT to +9\,nT across CS2 (10:11:41 -- 10:12:15 UT). In contrast to Events 1 and 2, the magnetic field does not dwell at a constant orientation in Region 3 (10:11:10 -- 10:11:41 UT), but instead shows large fluctuations. The total magnetic shear across these two current sheets is $95^\circ$. $v_l$ increases from -15\,km\,s$^{-1}$ to +20\,km\,s$^{-1}$, accompanied by a smaller increase in $|\mathbf{v_p^*}|$ from 10\,km\,s$^{-1}$ to 24\,km\,s$^{-1}$. The peak $|\mathbf{v_p^*}|$ of 35\,km\,s$^{-1}$ is observed in Region 3 and is roughly 75\% of the average local $v_A\sim46$\,km\,s$^{-1}$ in this region. We observe a small increase in $T_p$ from 14\,eV to 17\,eV, whereas $n_p$ decreases from a maximum of 30\,cm$^{-3}$ to 24\,cm$^{-3}$.

Figure \ref{fig6} shows the hodographs and Wal\'en plots for Event 3. Although not as distinct as Event 1, we still see an arc in the $lm$-plane hodographs and a quasi-vertical line in the $ln$-plane hodographs for all three current sheets. Across CS1 and CS2, the rotation in $\mathbf{B}$ is no longer clearly separated by a dwell period during which the field orientation remains roughly constant. This is caused by the magnetic field fluctuations in Region 3 causing $\mathbf{B}$ to 'double back' on itself in both hodographs. According to the $ln$-plane hodograph, $\mathbf{B}$ has a smaller $B_n$-component of $-0.1$\,nT than Events 1 and 2. The Wal\'en plot gradient of -0.297 for CS0 is below the range expected for an Alfv\'enic RD, whereas the gradients of +0.867 for CS1 and -0.697 for CS2 indicates that Alfv\'enic RDs are present across these two current sheets.

\section{Discussion}
\subsection{Evidence for reconnection at switchback boundaries}
Our overall findings suggest that the three observed events are magnetic switchbacks undergoing magnetic reconnection at their trailing edge boundaries. Based on the magnetic field observed in all three events, there is a polarity reversal in $B_R$, first at CS0 in each case and returning across CS1 and CS2 combined, consistent with magnetic switchbacks. For Event 1, we note the electron strahl PAD data supports this interpretation. As expected, the strahl pitch angle remains constant at $0^{\circ}$ both in the background HMF and in Region 2, the polarity-reversed section of the switchback. For Events 2 and 3, we use $v_{\alpha p}$ to confirm if these events are magnetic switchbacks \citep{Fedorov2021} as the electron strahl PAD data is unavailable. $v_{\alpha p}$ is positive in Regions 1, 3, and 4, where there are no polarity reversals in $B_R$. Conversely, $v_{\alpha p}$ is negative in Region 2 for both events; $v_{\alpha p}\sim-10$\,km\,s$^{-1}$ for Event 2 and $v_{\alpha p}\sim-20$\,km\,s$^{-1}$ for Event 3. This is in line with the expectation that $v_{\alpha p} > 0$ in the background solar wind and $v_{\alpha p} < 0$ inside the reversed section of a folded field configuration \citep{Marsch1982, Reisenfeld2001}.

Anti-correlation between the fluctuations in $\mathbf{B}$ and $\mathbf{v_p^*}$ in CS0 and Region 2 of Event 1 is consistent with an Alfv\'enic structure. However, the $|\mathbf{v_p^*}|$ enhancement of 27\,km\,s$^{-1}$ inside Region 2 is 68\% of the local Alfv\'en speed of 40\,km\,s$^{-1}$. This is less than the enhancements observed at switchbacks in the near-Sun solar wind, which are often roughly equivalent to the Alfv\'en speed \citep{Horbury2018, Kasper2019, Horbury2020a}. Combined with the decrease in $|\mathbf{B}|$, accompanying increase in $n_p$, and the Wal\'en plot for CS0 (Fig. \ref{fig2}e), these properties suggest that this event also has a non-Alfv\'enic component \citep{Kasper2019, Krasnoselskikh2020}. We do not observe similar correlations or any obvious change in $\mathbf{v_p^*}$ for Events 2 and 3. These velocity enhancements, if they exist, are considerably less than the local Alfv\'en speed. This property is also noted in reference to previously observed examples of reconnecting switchbacks \citep{Froment2021}.

The trailing edge boundary of all three switchbacks exhibit large increases in $|\mathbf{v_p^*}|$. The regions of accelerated flow at the trailing edge of the switchbacks are bound by a pair of current sheets CS1 and CS2 in each case, across which the fluctuations in $\mathbf{B}$ and $\mathbf{v_p^*}$ are anti-correlated on one side and correlated on the other. This bifurcation of the RCS at the trailing edge of the switchbacks and the presence of an accelerated outflow jet are consistent with the Gosling reconnection model \citep{Gosling2005a}. By contrast, the leading edge boundary of all three switchbacks show no signatures of current sheet bifurcation, and are instead comprised of a single current sheet CS0. Furthermore, with the exception of Event 1, no accelerated flows are observed across CS0 for the three events. This suggests that in each case, reconnection occurs only at the trailing edge boundary of the switchbacks, while the leading edge boundary of the switchback is non-reconnecting.

In the case of Event 1 (Fig. \ref{fig1}), the $\mathbf{v_p^*}$ enhancement in the polarity-reversed section of the switchback (Region 2) is oriented in the $+\mathbf{\hat{l}}$ direction, whereas the $\mathbf{v_p^*}$ enhancement in the trailing edge boundary reconnection outflow region (Region 3) is oriented in the $-\mathbf{\hat{l}}$ direction, suggesting that these two features are distinct from each other. The cause of the strahl dropout and broadening of the strahl PAD across CS1 is unknown but is not an instrumental effect, as a drop in the raw electron counts was also clearly detected by SWA-EAS at this time.

The hodographs show that five out of the six RCS have clear signatures associated with RDs, but the magnitudes of the line of best fit gradients for half of the Wal\'en plots fall below the $0.5$--$1$ range expected for an Alfv\'enic structure. This suggests that the reconnection outflow is sub-Alfv\'enic, a result that is not uncommon for reconnection in astrophysical plasmas \citep{Haggerty2018}. Modifying the Wal\'en relation (Eqn. \ref{eq3}) by factoring in a pressure anisotropy term \citep{Paschmann2008} makes no appreciable difference to the results of our analysis. Other reconnection models \citep{Petschek1964} and observational studies \citep{He2018, Phan2020} suggest that the reconnection outflow region boundaries can be composed of a combination of Alfv\'enic RDs and slow mode shocks. Shocks are not accounted for in the Wal\'en relation and may reduce the outflow velocity to sub-Alfv\'enic speeds \citep{Teh2009, Feng2017}. These may reduce the observed $\mathbf{v_p'}$ to $34$--$64\%$ of the predicted $v_A$ \citep{Phan2013, Phan2020}, which is more consistent with our Wal\'en plots. However, a detailed analysis of different reconnection models lies beyond the scope of our work.

\subsection{Switchback and reconnection geometry}
\begin{figure*}[t!]
   \centering
   \includegraphics[width=\textwidth]{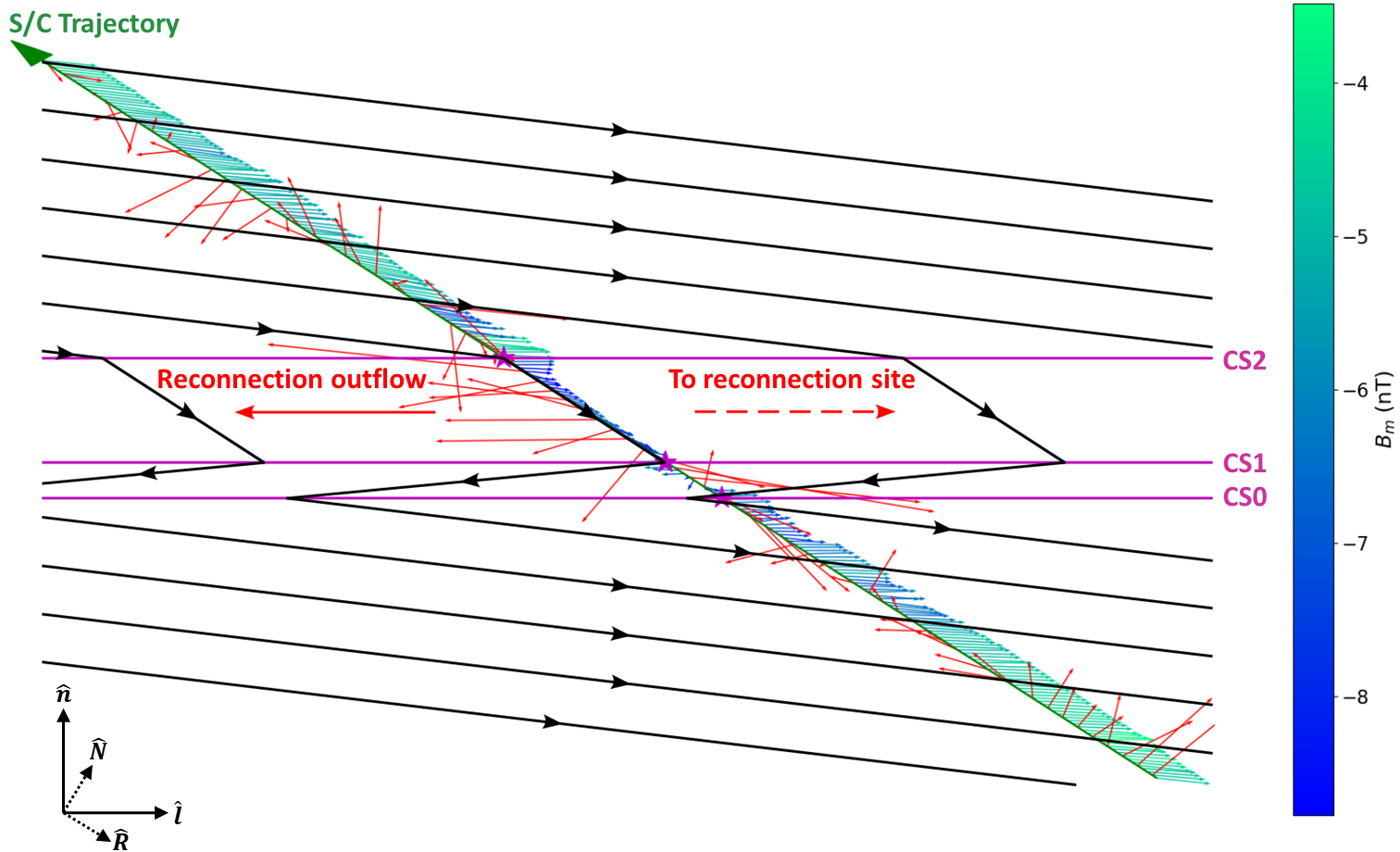}
   \caption{Feather plot of the $\mathbf{B}$ (blue/light green) and $\mathbf{v_p^*}$ vectors measured in Event 1 in the $ln$-plane with the spacecraft trajectory marked by the dark green arrow. The $B_m$ component of $\mathbf{B}$ is represented by the colour bar on the right. Overlaid on top is a possible interpretation of the magnetic field configuration of the switchback, shown here by the black arrows. The purple lines mark the assumed configuration of the current sheets CS0, CS1, and CS2. The purple stars show the locations where \textit{Solar Orbiter} crosses the current sheets.}
    \label{fig7}
\end{figure*}

Figure \ref{fig7} shows a feather plot of the magnetic field and proton velocity measurements recorded during Event 1. The measured $\mathbf{B}$ is shown by the blue/light green arrows and the measured $\mathbf{v_p^*}$ is shown by the solid red arrows. The colours of the $\mathbf{B}$ arrows represent the strength of the $B_m$ component of the magnetic field. We overlay a possible and consistent interpretation of the magnetic field configuration of the switchback on top, shown by the solid black arrows. As we are limited to measurements along the trajectory of \textit{Solar Orbiter} through this structure, the configuration shown here is one of many possible configurations that we consider consistent with the measurements. We assume that on large scales, the switchback is rigidly frozen into the bulk solar wind flow as it is convected across the spacecraft with constant velocity $\langle\mathbf{v_p}\rangle = (322.2, -5.6, -5.6)_{RTN}$\,km\,s$^{-1}$. Under this assumption, we map the measurement time stamps $t$ to spatial coordinates $\mathbf{r} = -(t-t_0)\langle\mathbf{v_p}\rangle$ where $t_0$ is an arbitrary reference time, defined here at 07:40:00 UT.

The dark green arrow represents the trajectory of \textit{Solar Orbiter} through Event 1, from the bottom right to the top left of the figure. We mark the locations where \textit{Solar Orbiter} crosses CS0, CS1, and CS2 with purple stars. In this assumed configuration, the spacecraft starts in the region of quiet anti-sunward ($+R$) HMF immediately preceding the switchback.  As the spacecraft crosses the leading edge boundary of the switchback (CS0), the polarity of the HMF reverses towards a sunward orientation and $|\mathbf{B}|$ decreases relative to the ambient HMF. $\mathbf{v_p^*}$ gradually increases and is directed in the $+\mathbf{\hat{l}}$ direction. 

The trailing edge boundary of the switchback, formed by the current sheets CS1 and CS2, together form a Gosling-type bifurcated RCS \citep{Gosling2005a} that bounds the reconnection outflow region. In order for the reconnection geometry to be consistent with the observed outflow, we require that the RCS extends back along the solid purple lines towards a reconnection site located off-page, in the $+\mathbf{\hat{l}}$ direction of the spacecraft trajectory. Inside the outflow region, $\mathbf{B}$ is roughly parallel with the spacecraft trajectory. Unlike at the leading edge of the switchback, $\mathbf{v_p^*}$ is directed in the $-l$ direction in this region. After crossing CS2, \textit{Solar Orbiter} exits the switchback and re-enters the surrounding solar wind, where conditions are similar to those observed immediately before the switchback encounter.

In the proposed scenario, magnetic reconnection occurs between oppositely directed field lines at the trailing edge boundary of the switchback. Within the overall geometry of the switchback, this topology may result in the formation of a magnetic flux rope on one side of the reconnection site and kinked magnetic field lines on the other. The strahl PAD in the outflow region (Region 3) shows that \textit{Solar Orbiter} passes through the side of the reconnection site containing open magnetic flux. Magnetic tension in these newly reconnected field causes them to recoil away from the reconnection site and straighten out, unwinding the switchback in the process. In this regard, our interpretation has many similarities with one proposed by \citet{Fedorov2021} to explain the formation of magnetic flux ropes at switchback-like structures observed near 1 au.

\subsection{Estimating the timescales for switchback erosion}
\begin{figure}
   \centering
   \includegraphics[width=\columnwidth]{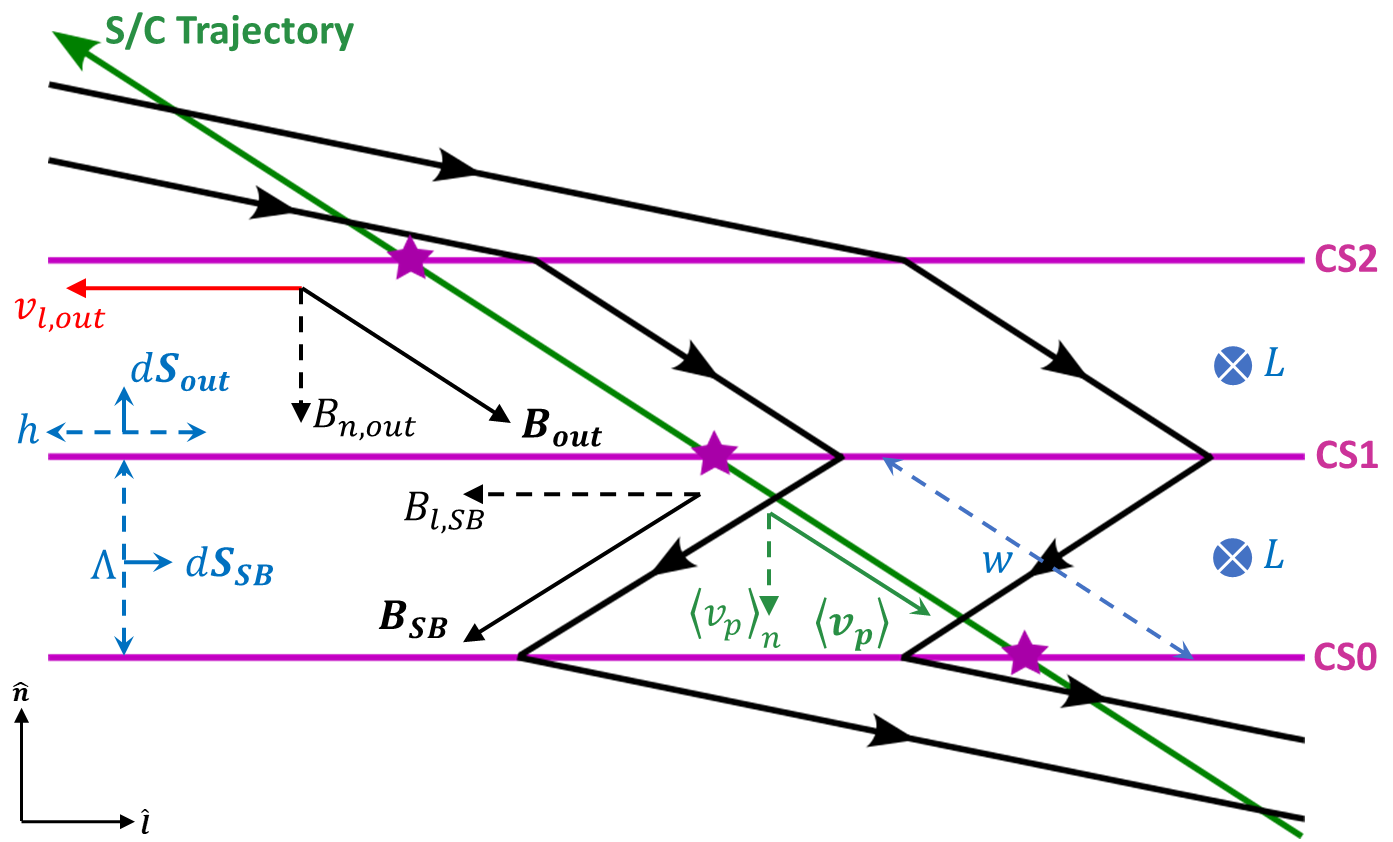}
    \caption{Simplified diagram of the switchback and reconnection geometry in Event 1, with quantities relevant to the calculation of $\tau$.}
    \label{fig8}
\end{figure}

\begin{table*}
    \caption{$\Lambda$: switchback width, $B_{SB}$: magnetic field strength in the switchback, $B_{out}$: magnetic field strength in the outflow region, $v_{out}$: exhaust outflow bulk speed, $\langle v_{SW}\rangle$: average solar wind speed, $\tau$: remaining lifetime of the switchback, $D$: convection distance travelled by switchback before fully eroding away.}                  
    \centering          
    \begin{tabular}{c c c c c c c c }     
        \hline\hline       
        Event & $\Lambda$ (km) & $B_{l,SB}$ (nT) & $B_{n,out}$ (nT) & $v_{l,out}$ (km s$^{-1}$) & $|\langle\mathbf{v_p}\rangle|$ (km s$^{-1}$) & $\tau$ (min) & $D$ (au) \\
        \hline                    
        Event 1 & 3570 & -4.8 & -1.0 & -7.2 & 322 & 40 & 0.005 \\
        Event 2 & 10100 & -9.7 & 0.5 & -28.3 & 439 & 126 & 0.02 \\
        Event 3 & 31700 & -7.5 & -0.1 & 15.4  & 443 & 2005 & 0.4 \\
        \hline                  
    \end{tabular}
    \label{tab4}
\end{table*}

We can estimate the remaining lifetime, $\tau$, of the three switchbacks discussed in this paper as they are being eroded by magnetic reconnection, assuming reconnection is the sole erosion mechanism and proceeds uniformly at the observed rate. This parameter depends on the magnetic flux $\phi_{SB}$ remaining in the polarity-reversed portion of the switchback which is yet to be reconnected, as well as the total rate of magnetic flux transport $2\dot{\phi}_{in}$ into the reconnection site from both sides of the reconnection region. As illustrated in Fig. \ref{fig7}, our proposed switchback geometry suggests that we only have direct measurements of the magnetic field and plasma in the outflow on one side of the reconnection region. These measurements allow us to quantify the rate of magnetic flux transport, $\dot{\phi}_{out}$, on that side of the outflow. Under the conservation of magnetic flux, $2\dot{\phi}_{in} = 2\dot{\phi}_{out}$, this leads to:
 
\begin{equation}
    \tau = \frac{\phi_{SB}}{\dot{\phi}_{in}} = \frac{\phi_{SB}}{\dot{\phi}_{out}}.
    \label{eq4}
\end{equation}

Figure \ref{fig8} shows a simplified diagram of the assumed switchback and reconnection geometry depicted in Fig. \ref{fig7}. We first consider the amount of magnetic flux $\phi_{out}$ transported by the reconnection outflow $\mathbf{v_{out}}$ in time $dt$. The general expression for magnetic flux through a surface composed of infinitesimal surface elements $d\mathbf{S}$ is given by $\phi=\int\mathbf{B}\cdot\mathbf{dS}$. In this 2D configuration, we define the surface element as $d\mathbf{S_{out}} = hL\mathbf{\hat{n}}$, where $h = v_{l,out}dt$ is the distance the reconnected field lines are convected by the outflow in time $t$, and $L$ is the out-of-plane extent of the switchback. Hence, 
\begin{equation}
    \phi_{out}=\int\mathbf{B_{out}}\cdot\mathbf{dS_{out}} \approx B_{n,out}v_{l,out}Ldt,
    \label{eq5}
\end{equation}   
 where $B_{n,out}$ is the average $B_n$-component of the magnetic field in the outflow region. This is equivalent to
 
\begin{equation}
    \dot{\phi}_{out} \approx B_{n,out}v_{l,out}L.
    \label{eq6}
\end{equation} 

The distance $w$ travelled by \textit{Solar Orbiter} in the polarity-reversed section of the switchback (Region 2) is trajectory-dependent and hence, is an unreliable measure for the switchback width. We instead use $\Lambda = \langle\mathbf{v_p}\rangle_n dt_{SB}$, the perpendicular distance between CS0 and CS1, to estimate the width of the polarity-reversed section of the switchback. Here, $dt_{SB}$ is the crossing duration of Region 2. Applying similar reasoning to the derivation of $\phi_{out}$ above, we estimate the $\phi_{SB}$ to be:
\begin{equation}
    \phi_{SB}=\int\mathbf{B_{SB}}\cdot\mathbf{dS_{SB}} \approx B_{l,SB}\Lambda L.
    \label{eq7}
\end{equation}   
Here, we have oriented the surface element $d\mathbf{S_{SB}} = \Lambda L\mathbf{\hat{l}}$ along the $l$-direction, as $B_{l,SB}$ is the component of $\mathbf{B_{SB}}$ that reconnects. Finally, we substitute Eqs. \ref{eq6} and \ref{eq7} into Eq. \ref{eq4} to obtain the time remaining until complete erosion of the switchback:
\begin{equation}
    \tau = \frac{B_{l,SB}\Lambda}{B_{n,out}v_{l,out}}
    \label{eq8}
\end{equation}
We estimate the remaining convection distance $D$ until the complete erosion of the switchback as $D\simeq|\langle\mathbf{v_p}\rangle|\tau$.

Table \ref{tab4} shows the estimated $\tau$ and $D$ for the three events discussed in this paper. From Eq. \ref{eq8}, $\tau$ depends linearly on switchback width $\Lambda$, which determines the amount of magnetic flux remaining in the polarity-reversed section of the switchback; and is inversely proportional to $v_{l,out}$, which indicates the rate at which reconnected flux is transported away from the reconnection site. Since Event 1 has the smallest width of $\Lambda = 3570$\,km and the largest absolute $B_{n,out}$ of 1.0\,nT, it has the shortest $\tau$ of 40 minutes despite having the slowest absolute $v_{l,out}$ of 7.2\,km\,s$^{-1}$. Given the small $\Lambda$ and short $\tau$ compared to the other two events, this suggests that Event 1 may be a switchback that has almost been completely eroded by reconnection. Conversely, Event 3 is the widest with $\Lambda = 31700$\,km and the smallest $B_{n,out}$ of 0.1\,nT, a factor of ten smaller than $B_{n,out}$ for Event 1. As a result, it has the longest $\tau = 2005$ minutes out of the three events. Event 2 is roughly three times wider than Event 1 with $\Lambda = 10100$\,km and has $B_{l,SB} = 9.7$\,nT twice as large as Event 1, but has the greatest $v_{l,out} = 28.3$\,km\,s$^{-1}$. Its $\tau = 126$ minutes is thrice as long as for Event 1. $D$ travelled by these three switchbacks before they fully erode range from 0.005 au (Event 1)\,to 0.4 au\,(Event 3).

\subsection{Implications on switchback formation and evolution in the heliosphere}
A key assumption we have made in our calculations in Section 4.3 is that reconnection proceeds uniformly at the observed rate. Because we have no information about the time history of these switchbacks as they evolve from their place of origin to their place of detection, we do not know when or where the onset of reconnection occurs. Therefore, neither $\tau$ nor $D$ should be taken as the actual time or distance between reconnection onset and complete erosion of the switchback. 

However, $\tau$ and $D$ are both small compared to the characteristic timescales and distances of the solar wind expansion, which suggests that reconnection is a fast and efficient mechanism through which switchbacks can be eroded. To highlight this point, let us assume that the onset of reconnection occurs at heliocentric distances similar to PSP perihelion 1 ($\sim0.2$\,au), during which PSP made its observations of prominent switchbacks and switchback patches \citep{Bale2019, Kasper2019, Horbury2020a}. If the reconnection rate remains constant during transport in the solar wind, $\Lambda$ for the observed switchbacks at these distances would be 0.1 -- 0.5 solar radii. This is significantly larger than what was observed by PSP and has two possible implications.

The first is that the switchbacks are formed near the Sun and propagate stably into interplanetary space, before encountering conditions enabling the onset of reconnection and thus the rapid erosion of the switchback. This scenario would explain the rarity of observations of reconnection at switchback boundaries, as the observing spacecraft would need to serendipitously encounter the switchback at almost the same time as reconnection onset. It would also explain why fewer switchbacks are observed at 0.6 -- 0.7\,au by \textit{Solar Orbiter} compared to PSP at heliocentric distances $<0.2$\,au. Furthermore, \citet{Tenerani2020} demonstrate that large switchbacks formed in the corona can only survive out to $\sim0.2$\,au if the background solar wind conditions are sufficiently calm, before the parametric decay instability causes them to decay.

The second explanation is that the switchbacks are formed in-situ in the solar wind at a time much closer to the time of their detection. This is supported by new results from \citet{Macneil2020} and \citet{Pecora2022} that suggest the occurrence rate of magnetic switchbacks increases with heliocentric distance. There is the possibility that two (or more) populations of switchbacks exist: those that form in the Sun's corona, and those that form in the solar wind \citep{Tenerani2021}.

\section{Conclusions}
Using \textit{Solar Orbiter} data from 10 August and 30 August 2021, we identify three magnetic switchbacks at heliocentric distances between 0.6 -- 0.7 au. The trailing edge boundaries of all three events show signatures of jetting and current sheet bifurcation consistent with the Gosling reconnection model \citep{Gosling2005a}. 

We propose a possible configuration of the switchback observed on 10 August and reconnection geometry based on measurements of the switchback. In this scenario, reconnection at the trailing edge boundary of the switchback results in the formation of a magnetic flux rope on one side of the reconnection site and kinked field lines on the other. Magnetic tension causes the reconnected field lines to recoil away from the reconnection site, resulting in the unwinding of the switchback. In this paper, we only find cases in which magnetic reconnection occurs at the trailing edge boundary of switchbacks. However, in principle, this process may also occur at the leading edge boundary of switchbacks or at the leading and trailing edge boundaries simultaneously. Although magnetic tension acts naturally to straighten field line kinks in non-reconnecting switchbacks as well, our observation-driven scenario suggests that reconnection can increase the rate at which these structures unwind.

Our estimates of the remaining lifetime of the switchbacks suggest that they erode within a few minutes to a few hours after being observed by \textit{Solar Orbiter}. During this time, the switchbacks are carried a further 0.005 -- 0.4 au by the surrounding solar wind flow. If typical, these results could explain why switchbacks are rarely seen at 1 au and has implications on how these structures form and evolve in the heliosphere. The short $\tau$ and small $D$ relative to the characteristic timescales and distances of the solar wind expansion show that reconnection is an efficient process for switchback erosion. This suggests that the onset of reconnection must occur during transport in the solar wind in our examples and supports theories of switchback formation in-situ in the solar wind.

There are some caveats to our results and interpretation. The use of single-spacecraft measurements limits our knowledge of the magnetic field and solar wind conditions inside the switchback to what is observed along the spacecraft's trajectory. Furthermore, we also have no information about the time history of the switchbacks. Consequently, we do not know when the onset of reconnection occurs at the switchback boundaries, nor whether this process creates a flux rope embedded within the switchback. Therefore, our interpretation must be understood as one possible scenario that we consider with the measured knowledge of the field and plasma geometry. 

In order to further develop the ideas presented here, multi-spacecraft observations will be needed. Radial line-up opportunities between \textit{Solar Orbiter} and other spacecraft such as PSP will allow us to track the temporal evolution of individual switchbacks with heliocentric distance, and identify the conditions required for reconnection to occur at their boundaries. Repeating our analysis on PSP events \citep{Froment2021} and comparing the results with the ones discussed here would also be an interesting idea to explore in future studies.

Our model predicts that reconnection will convert a portion of the switchback into a magnetic flux rope disconnected from the Sun. Such a structure will appear as a reversal in the HMF polarity but can be distinguished from a switchback in the strahl PAD data. Simultaneous multi-point measurements of the switchback, reconnection outflow region, and flux rope by constellation-type missions such as \textit{Cluster} \citep{Escoubet1997}, \textit{MMS} \citep{Burch2016}, and the upcoming \textit{HelioSwarm} \citep{Klein2019, Broeren2021, Matthaeus2022} will allow us to verify the validity of our model. These types of measurements can also better constrain the 3D geometry of these structures and isolate spatial variations from temporal variations.

\begin{acknowledgements}
\textit{Solar Orbiter} is a space mission of international collaboration between ESA and NASA, operated by ESA. Solar Orbiter Solar Wind Analyser (SWA) data are derived from scientific sensors which have been designed and created, and are operated under funding provided in numerous contracts 
from the UK Space Agency (UKSA), the UK Science and Technology Facilities Council (STFC), the Agenzia Spaziale Italiana (ASI), the Centre National d’Etudes Spatiales (CNES, France), the Centre National de la Recherche Scientifique (CNRS, France), the Czech contribution to the ESA PRODEX programme and NASA. Solar Orbiter SWA work at UCL/MSSL was funded by the UK Space Agency under STFC grants ST/T001356/1, ST/S000240/1, ST/X002152/1, ST/W001004/1 and ST/P003826/1. The Solar Orbiter magnetometer was funded by the UK Space Agency (grant ST/X002098/1). T.S.H is supported by STFC grant ST/S000364/1. The author would also like to thank the anonymous referee for their detailed and constructive feedback on an earlier version of the manuscript. For the purpose of open access, the author has applied a Creative Commons Attribution (CC BY) licence to any Author Accepted Manuscript version arising.
\end{acknowledgements}

\bibliographystyle{aa}
\bibliography{ref}

\begin{thebibliography}{76}
\expandafter\ifx\csname natexlab\endcsname\relax\def\natexlab#1{#1}\fi

\bibitem[{{Adhikari} {et~al.}(2019){Adhikari}, {Khabarova}, {Zank}, \&
  {Zhao}}]{Adhikari2019}
{Adhikari}, L., {Khabarova}, O., {Zank}, G.~P., \& {Zhao}, L.-L. 2019, \apj,
  873, 72

\bibitem[{{Bale} {et~al.}(2019){Bale}, {Badman}, {Bonnell}, {Bowen}, {Burgess},
  {Case}, {Cattell}, {Chandran}, {Chaston}, {Chen}, {Drake}, {de Wit},
  {Eastwood}, {Ergun}, {Farrell}, {Fong}, {Goetz}, {Goldstein}, {Goodrich},
  {Harvey}, {Horbury}, {Howes}, {Kasper}, {Kellogg}, {Klimchuk}, {Korreck},
  {Krasnoselskikh}, {Krucker}, {Laker}, {Larson}, {MacDowall}, {Maksimovic},
  {Malaspina}, {Martinez-Oliveros}, {McComas}, {Meyer-Vernet}, {Moncuquet},
  {Mozer}, {Phan}, {Pulupa}, {Raouafi}, {Salem}, {Stansby}, {Stevens}, {Szabo},
  {Velli}, {Woolley}, \& {Wygant}}]{Bale2019}
{Bale}, S.~D., {Badman}, S.~T., {Bonnell}, J.~W., {et~al.} 2019, \nat, 576, 237

\bibitem[{{Balogh} {et~al.}(1999){Balogh}, {Forsyth}, {Lucek}, {Horbury}, \&
  {Smith}}]{Balogh1999}
{Balogh}, A., {Forsyth}, R.~J., {Lucek}, E.~A., {Horbury}, T.~S., \& {Smith},
  E.~J. 1999, \grl, 26, 631

\bibitem[{{Broeren} {et~al.}(2021){Broeren}, {Klein}, {TenBarge}, {Dors},
  {Roberts}, \& {Verscharen}}]{Broeren2021}
{Broeren}, T., {Klein}, K.~G., {TenBarge}, J.~M., {et~al.} 2021, Frontiers in
  Astronomy and Space Sciences, 8, 144

\bibitem[{{Burch} {et~al.}(2016){Burch}, {Moore}, {Torbert}, \&
  {Giles}}]{Burch2016}
{Burch}, J.~L., {Moore}, T.~E., {Torbert}, R.~B., \& {Giles}, B.~L. 2016, \ssr,
  199, 5

\bibitem[{{Cassak}(2016)}]{Cassak2016}
{Cassak}, P.~A. 2016, Space Weather, 14, 186

\bibitem[{{de Hoffmann} \& {Teller}(1950)}]{deHoffmann1950}
{de Hoffmann}, F. \& {Teller}, E. 1950, Physical Review, 80, 692

\bibitem[{{De Marco} {et~al.}(2023){De Marco}, {Bruno}, {Jagarlamudi},
  {D'Amicis}, {Marcucci}, {Fortunato}, {Perrone}, {Telloni}, {Owen}, {Louarn},
  {Fedorov}, {Livi}, \& {Horbury}}]{DeMarco2023}
{De Marco}, R., {Bruno}, R., {Jagarlamudi}, V.~K., {et~al.} 2023, \aap, 669,
  A108

\bibitem[{{de Pablos} {et~al.}(2022){de Pablos}, {Samanta}, {Badman},
  {Schwanitz}, {Bahauddin}, {Harra}, {Petrie}, {Mac Cormack}, {Mandrini},
  {Raouafi}, {Martinez Pillet}, \& {Velli}}]{DePablos2022}
{de Pablos}, D., {Samanta}, T., {Badman}, S.~T., {et~al.} 2022, \solphys, 297,
  90

\bibitem[{{Dong} {et~al.}(2017){Dong}, {Dunlop}, {Trattner}, {Phan}, {Fu},
  {Cao}, {Russell}, {Giles}, {Torbert}, {Le}, \& {Burch}}]{Dong2017}
{Dong}, X.-C., {Dunlop}, M.~W., {Trattner}, K.~J., {et~al.} 2017, \grl, 44,
  5951

\bibitem[{{Drake} {et~al.}(2021){Drake}, {Agapitov}, {Swisdak}, {Badman},
  {Bale}, {Horbury}, {Kasper}, {MacDowall}, {Mozer}, {Phan}, {Pulupa}, {Szabo},
  \& {Velli}}]{Drake2021}
{Drake}, J.~F., {Agapitov}, O., {Swisdak}, M., {et~al.} 2021, \aap, 650, A2

\bibitem[{{Drake} {et~al.}(2009){Drake}, {Cassak}, {Shay}, {Swisdak}, \&
  {Quataert}}]{Drake2009}
{Drake}, J.~F., {Cassak}, P.~A., {Shay}, M.~A., {Swisdak}, M., \& {Quataert},
  E. 2009, \apj, 700, L16

\bibitem[{{Dudok de Wit} {et~al.}(2020){Dudok de Wit}, {Krasnoselskikh},
  {Bale}, {Bonnell}, {Bowen}, {Chen}, {Froment}, {Goetz}, {Harvey},
  {Jagarlamudi}, {Larosa}, {MacDowall}, {Malaspina}, {Matthaeus}, {Pulupa},
  {Velli}, \& {Whittlesey}}]{DudokDeWit2020}
{Dudok de Wit}, T., {Krasnoselskikh}, V.~V., {Bale}, S.~D., {et~al.} 2020,
  \apjs, 246, 39

\bibitem[{{En{\v{z}}l} {et~al.}(2014){En{\v{z}}l}, {P{\v{r}}ech},
  {{\v{S}}afr{\'{a}}nkov{\'{a}}}, \& {N{\v{e}}me{\v{c}}ek}}]{Enzl2014}
{En{\v{z}}l}, J., {P{\v{r}}ech}, L., {{\v{S}}afr{\'{a}}nkov{\'{a}}}, J., \&
  {N{\v{e}}me{\v{c}}ek}, Z. 2014, \apj, 796, 21

\bibitem[{{Escoubet} {et~al.}(1997){Escoubet}, {Schmidt}, \&
  {Goldstein}}]{Escoubet1997}
{Escoubet}, C.~P., {Schmidt}, R., \& {Goldstein}, M.~L. 1997, \ssr, 79, 11

\bibitem[{{Fargette} {et~al.}(2021){Fargette}, {Lavraud}, {Rouillard},
  {R{\'e}ville}, {Dudok De Wit}, {Froment}, {Halekas}, {Phan}, {Malaspina},
  {Bale}, {Kasper}, {Louarn}, {Case}, {Korreck}, {Larson}, {Pulupa}, {Stevens},
  {Whittlesey}, \& {Berthomier}}]{Fargette2021}
{Fargette}, N., {Lavraud}, B., {Rouillard}, A.~P., {et~al.} 2021, \apj, 919, 96

\bibitem[{{Fedorov} {et~al.}(2021){Fedorov}, {Louarn}, {Owen}, {Horbury},
  {Prech}, {Durovcova}, {Barthe}, {Rouillard}, {Kasper}, {Bale}, {Bruno},
  {O'Brien}, {Evans}, {Angelini}, {Larson}, {Livi}, {Lavraud}, {Andre},
  {Genot}, {Penou}, {Mele}, \& {Fortunato}}]{Fedorov2021}
{Fedorov}, A., {Louarn}, P., {Owen}, C.~J., {et~al.} 2021, \aap, 656, A40

\bibitem[{{Feldman} {et~al.}(1975){Feldman}, {Asbridge}, {Bame}, {Montgomery},
  \& {Gary}}]{Feldman1975}
{Feldman}, W.~C., {Asbridge}, J.~R., {Bame}, S.~J., {Montgomery}, M.~D., \&
  {Gary}, S.~P. 1975, \jgr, 80, 4181

\bibitem[{{Feng} {et~al.}(2017){Feng}, {Li}, {Wang}, \& {Zhao}}]{Feng2017}
{Feng}, H., {Li}, Q., {Wang}, J., \& {Zhao}, G. 2017, \solphys, 292, 53

\bibitem[{{Fisk} \& {Kasper}(2020)}]{Fisk2020}
{Fisk}, L.~A. \& {Kasper}, J.~C. 2020, \apjl, 894, L4

\bibitem[{{Froment} {et~al.}(2021){Froment}, {Krasnoselskikh}, {Dudok de Wit},
  {Agapitov}, {Fargette}, {Lavraud}, {Larosa}, {Kretzschmar}, {Jagarlamudi},
  {Velli}, {Malaspina}, {Whittlesey}, {Bale}, {Case}, {Goetz}, {Kasper},
  {Korreck}, {Larson}, {MacDowall}, {Mozer}, {Pulupa}, {Revillet}, \&
  {Stevens}}]{Froment2021}
{Froment}, C., {Krasnoselskikh}, V., {Dudok de Wit}, T., {et~al.} 2021, \aap,
  650, A5

\bibitem[{{Gosling}(2012)}]{Gosling2012}
{Gosling}, J.~T. 2012, \ssr, 172, 187

\bibitem[{{Gosling} {et~al.}(2006{\natexlab{a}}){Gosling}, {Eriksson}, \&
  {Schwenn}}]{Gosling2006b}
{Gosling}, J.~T., {Eriksson}, S., \& {Schwenn}, R. 2006{\natexlab{a}}, Journal
  of Geophysical Research (Space Physics), 111, A10102

\bibitem[{{Gosling} {et~al.}(2006{\natexlab{b}}){Gosling}, {McComas}, {Skoug},
  \& {Smith}}]{Gosling2006a}
{Gosling}, J.~T., {McComas}, D.~J., {Skoug}, R.~M., \& {Smith}, C.~W.
  2006{\natexlab{b}}, \grl, 33, L17102

\bibitem[{{Gosling} \& {Phan}(2013)}]{Gosling2013}
{Gosling}, J.~T. \& {Phan}, T.~D. 2013, \apj, 763, L39

\bibitem[{{Gosling} {et~al.}(2007){Gosling}, {Phan}, {Lin}, \&
  {Szabo}}]{Gosling2007}
{Gosling}, J.~T., {Phan}, T.~D., {Lin}, R.~P., \& {Szabo}, A. 2007, \grl, 34,
  L15110

\bibitem[{{Gosling} {et~al.}(2005{\natexlab{a}}){Gosling}, {Skoug}, {McComas},
  \& {Smith}}]{Gosling2005a}
{Gosling}, J.~T., {Skoug}, R.~M., {McComas}, D.~J., \& {Smith}, C.~W.
  2005{\natexlab{a}}, \jgr {} Space Physics, 110

\bibitem[{{Gosling} {et~al.}(2005{\natexlab{b}}){Gosling}, {Skoug}, {McComas},
  \& {Smith}}]{Gosling2005b}
{Gosling}, J.~T., {Skoug}, R.~M., {McComas}, D.~J., \& {Smith}, C.~W.
  2005{\natexlab{b}}, \grl, 32, L05105

\bibitem[{{Haggerty} {et~al.}(2018){Haggerty}, {Shay}, {Chasapis}, {Phan},
  {Drake}, {Malakit}, {Cassak}, \& {Kieokaew}}]{Haggerty2018}
{Haggerty}, C.~C., {Shay}, M.~A., {Chasapis}, A., {et~al.} 2018, PhPl, 25,
  102120

\bibitem[{{He} {et~al.}(2018){He}, {Zhu}, {Chen}, {Salem}, {Stevens}, {Li},
  {Ruan}, {Zhang}, \& {Tu}}]{He2018}
{He}, J., {Zhu}, X., {Chen}, Y., {et~al.} 2018, \apj, 856, 148

\bibitem[{{Hesse} \& {Cassak}(2020)}]{Hesse2020}
{Hesse}, M. \& {Cassak}, P.~A. 2020, Journal of Geophysical Research (Space
  Physics), 125, e25935

\bibitem[{{Horbury} {et~al.}(2018){Horbury}, {Matteini}, \&
  {Stansby}}]{Horbury2018}
{Horbury}, T.~S., {Matteini}, L., \& {Stansby}, D. 2018, \mnras, 478, 1980

\bibitem[{{Horbury} {et~al.}(2020{\natexlab{a}}){Horbury}, {O'Brien}, {Carrasco
  Blazquez}, {Bendyk}, {Brown}, {Hudson}, {Evans}, {Oddy}, {Carr}, {Beek},
  {Cupido}, {Bhattacharya}, {Dominguez}, {Matthews}, {Myklebust}, {Whiteside},
  {Bale}, {Baumjohann}, {Burgess}, {Carbone}, {Cargill}, {Eastwood},
  {Erd{\"o}s}, {Fletcher}, {Forsyth}, {Giacalone}, {Glassmeier}, {Goldstein},
  {Hoeksema}, {Lockwood}, {Magnes}, {Maksimovic}, {Marsch}, {Matthaeus},
  {Murphy}, {Nakariakov}, {Owen}, {Owens}, {Rodriguez-Pacheco}, {Richter},
  {Riley}, {Russell}, {Schwartz}, {Vainio}, {Velli}, {Vennerstrom}, {Walsh},
  {Wimmer-Schweingruber}, {Zank}, {M{\"u}ller}, {Zouganelis}, \&
  {Walsh}}]{Horbury2020b}
{Horbury}, T.~S., {O'Brien}, H., {Carrasco Blazquez}, I., {et~al.}
  2020{\natexlab{a}}, \aap, 642, A9

\bibitem[{{Horbury} {et~al.}(2020{\natexlab{b}}){Horbury}, {Woolley}, {Laker},
  {Matteini}, {Eastwood}, {Bale}, {Velli}, {Chandran}, {Phan}, {Raouafi},
  {Goetz}, {Harvey}, {Pulupa}, {Klein}, {Dudok de Wit}, {Kasper}, {Korreck},
  {Case}, {Stevens}, {Whittlesey}, {Larson}, {MacDowall}, {Malaspina}, \&
  {Livi}}]{Horbury2020a}
{Horbury}, T.~S., {Woolley}, T., {Laker}, R., {et~al.} 2020{\natexlab{b}},
  \apjs, 246, 45

\bibitem[{{Hudson}(1970)}]{Hudson1970}
{Hudson}, P.~D. 1970, \planss, 18, 1611

\bibitem[{{Johnston} {et~al.}(2022){Johnston}, {Squire}, {Mallet}, \&
  {Meyrand}}]{Johnston2022}
{Johnston}, Z., {Squire}, J., {Mallet}, A., \& {Meyrand}, R. 2022, Physics of
  Plasmas, 29, 072902

\bibitem[{{Kasper} {et~al.}(2019){Kasper}, {Bale}, {Belcher}, {Berthomier},
  {Case}, {Chandran}, {Curtis}, {Gallagher}, {Gary}, {Golub}, {Halekas}, {Ho},
  {Horbury}, {Hu}, {Huang}, {Klein}, {Korreck}, {Larson}, {Livi}, {Maruca},
  {Lavraud}, {Louarn}, {Maksimovic}, {Martinovic}, {McGinnis}, {Pogorelov},
  {Richardson}, {Skoug}, {Steinberg}, {Stevens}, {Szabo}, {Velli},
  {Whittlesey}, {Wright}, {Zank}, {MacDowall}, {McComas}, {McNutt}, {Pulupa},
  {Raouafi}, \& {Schwadron}}]{Kasper2019}
{Kasper}, J.~C., {Bale}, S.~D., {Belcher}, J.~W., {et~al.} 2019, \nat, 576, 228

\bibitem[{{Khabarova} {et~al.}(2015){Khabarova}, {Zank}, {Li}, {le Roux},
  {Webb}, {Dosch}, \& {Malandraki}}]{Khabarova2015}
{Khabarova}, O., {Zank}, G.~P., {Li}, G., {et~al.} 2015, \apj, 808, 181

\bibitem[{{Khrabrov} \& {Sonnerup}(1998)}]{Khrabrov1998}
{Khrabrov}, A.~V. \& {Sonnerup}, B.~U.~{\"O}. 1998, ISSI Scientific Reports
  Series, 1, 221

\bibitem[{{Kieokaew} {et~al.}(2021){Kieokaew}, {Lavraud}, {Yang}, {Matthaeus},
  {Ruffolo}, {Stawarz}, {Aizawa}, {Foullon}, {G{\'e}not}, {Pinto}, {Fargette},
  {Louarn}, {Rouillard}, {Fedorov}, {Penou}, {Owen}, {Horbury}, {O'Brien},
  {Evans}, \& {Angelini}}]{Kieokaew2021}
{Kieokaew}, R., {Lavraud}, B., {Yang}, Y., {et~al.} 2021, \aap, 656, A12

\bibitem[{{Klein} {et~al.}(2019){Klein}, {Alexandrova}, {Bookbinder},
  {Caprioli}, {Case}, {Chandran}, {Chen}, {Horbury}, {Jian}, {Kasper}, {Le
  Contel}, {Maruca}, {Matthaeus}, {Retino}, {Roberts}, {Schekochihin}, {Skoug},
  {Smith}, {Steinberg}, {Spence}, {Vasquez}, {TenBarge}, {Verscharen}, \&
  {Whittlesey}}]{Klein2019}
{Klein}, K.~G., {Alexandrova}, O., {Bookbinder}, J., {et~al.} 2019, arXiv
  e-prints, arXiv:1903.05740

\bibitem[{{Krasnoselskikh} {et~al.}(2020){Krasnoselskikh}, {Larosa},
  {Agapitov}, {de Wit}, {Moncuquet}, {Mozer}, {Stevens}, {Bale}, {Bonnell},
  {Froment}, {Goetz}, {Goodrich}, {Harvey}, {Kasper}, {MacDowall}, {Malaspina},
  {Pulupa}, {Raouafi}, {Revillet}, {Velli}, \& {Wygant}}]{Krasnoselskikh2020}
{Krasnoselskikh}, V., {Larosa}, A., {Agapitov}, O., {et~al.} 2020, \apj, 893,
  93

\bibitem[{{Macneil} {et~al.}(2020){Macneil}, {Owens}, {Wicks}, {Lockwood},
  {Bentley}, \& {Lang}}]{Macneil2020}
{Macneil}, A.~R., {Owens}, M.~J., {Wicks}, R.~T., {et~al.} 2020, \mnras, 494,
  3642

\bibitem[{{Marsch} {et~al.}(1982){Marsch}, {Rosenbauer}, {Schwenn},
  {Muehlhaeuser}, \& {Neubauer}}]{Marsch1982}
{Marsch}, E., {Rosenbauer}, H., {Schwenn}, R., {Muehlhaeuser}, K.~H., \&
  {Neubauer}, F.~M. 1982, \jgr, 87, 35

\bibitem[{{Matteini} {et~al.}(2014){Matteini}, {Horbury}, {Neugebauer}, \&
  {Goldstein}}]{Matteini2014}
{Matteini}, L., {Horbury}, T.~S., {Neugebauer}, M., \& {Goldstein}, B.~E. 2014,
  \grl, 41, 259

\bibitem[{{Matthaeus} {et~al.}(2022){Matthaeus}, {Adhikari}, {Bandyopadhyay},
  {Brown}, {Bruno}, {Borovsky}, {Carbone}, {Caprioli}, {Chasapis}, {Chhiber},
  {Dasso}, {Dmitruk}, {Del Zanna}, {Dmitruk}, {Franci}, {Gary}, {Goldstein},
  {Gomez}, {Greco}, {Horbury}, {Ji}, {Kasper}, {Klein}, {Landi}, {Li},
  {Malara}, {Maruca}, {Mininni}, {Oughton}, {Papini}, {Parashar}, {Pecora},
  {Petrosyan}, {Pouquet}, {Retino}, {Roberts}, {Ruffolo}, {Servidio}, {Spence},
  {Smith}, {Stawarz}, {TenBarge}, {Vasquez}, {Vaivads}, {Valentini}, {Velli},
  {Verdini}, {Verscharen}, {Whittlesey}, {Wicks}, {Yang}, \&
  {Zimbardo}}]{Matthaeus2022}
{Matthaeus}, W.~H., {Adhikari}, S., {Bandyopadhyay}, R., {et~al.} 2022, arXiv
  e-prints, arXiv:2211.12676

\bibitem[{{McComas} {et~al.}(1994){McComas}, {Gosling}, {Hammond}, {Moldwin},
  {Phillips}, \& {Forsyth}}]{McComas1994}
{McComas}, D.~J., {Gosling}, J.~T., {Hammond}, C.~M., {et~al.} 1994, \grl, 21,
  1751

\bibitem[{{Mistry} {et~al.}(2017){Mistry}, {Eastwood}, {Phan}, \&
  {Hietala}}]{Mistry2017}
{Mistry}, R., {Eastwood}, J.~P., {Phan}, T.~D., \& {Hietala}, H. 2017, \jgr {}
  Space Physics, 122, 5895

\bibitem[{{Neugebauer} \& {Sterling}(2021)}]{Neugebauer2021}
{Neugebauer}, M. \& {Sterling}, A.~C. 2021, \apjl, 920, L31

\bibitem[{{Owen} {et~al.}(2020){Owen}, {Bruno}, {Livi}, {Louarn}, {Al Janabi},
  {Allegrini}, {Amoros}, {Baruah}, {Barthe}, {Berthomier}, {Bordon},
  {Brockley-Blatt}, {Brysbaert}, {Capuano}, {Collier}, {DeMarco}, {Fedorov},
  {Ford}, {Fortunato}, {Fratter}, {Galvin}, {Hancock}, {Heirtzler}, {Kataria},
  {Kistler}, {Lepri}, {Lewis}, {Loeffler}, {Marty}, {Mathon}, {Mayall}, {Mele},
  {Ogasawara}, {Orlandi}, {Pacros}, {Penou}, {Persyn}, {Petiot}, {Phillips},
  {P{\v{r}}ech}, {Raines}, {Reden}, {Rouillard}, {Rousseau}, {Rubiella},
  {Seran}, {Spencer}, {Thomas}, {Trevino}, {Verscharen}, {Wurz}, {Alapide},
  {Amoruso}, {Andr{\'e}}, {Anekallu}, {Arciuli}, {Arnett}, {Ascolese},
  {Bancroft}, {Bland}, {Brysch}, {Calvanese}, {Castronuovo},
  {{\v{C}}erm{\'a}k}, {Chornay}, {Clemens}, {Coker}, {Collinson}, {D'Amicis},
  {Dandouras}, {Darnley}, {Davies}, {Davison}, {De Los Santos}, {Devoto},
  {Dirks}, {Edlund}, {Fazakerley}, {Ferris}, {Frost}, {Fruit}, {Garat},
  {G{\'e}not}, {Gibson}, {Gilbert}, {de Giosa}, {Gradone}, {Hailey}, {Horbury},
  {Hunt}, {Jacquey}, {Johnson}, {Lavraud}, {Lawrenson}, {Leblanc}, {Lockhart},
  {Maksimovic}, {Malpus}, {Marcucci}, {Mazelle}, {Monti}, {Myers}, {Nguyen},
  {Rodriguez-Pacheco}, {Phillips}, {Popecki}, {Rees}, {Rogacki}, {Ruane},
  {Rust}, {Salatti}, {Sauvaud}, {Stakhiv}, {Stange}, {Stubbs}, {Taylor},
  {Techer}, {Terrier}, {Thibodeaux}, {Urdiales}, {Varsani}, {Walsh}, {Watson},
  {Wheeler}, {Willis}, {Wimmer-Schweingruber}, {Winter}, {Yardley}, \&
  {Zouganelis}}]{Owen2020}
{Owen}, C.~J., {Bruno}, R., {Livi}, S., {et~al.} 2020, \aap, 642, A16

\bibitem[{{Owens} {et~al.}(2013){Owens}, {Crooker}, \& {Lockwood}}]{Owens2013a}
{Owens}, M.~J., {Crooker}, N.~U., \& {Lockwood}, M. 2013, Journal of
  Geophysical Research (Space Physics), 118, 1868

\bibitem[{{Owens} \& {Forsyth}(2013)}]{Owens2013b}
{Owens}, M.~J. \& {Forsyth}, R.~J. 2013, Living Reviews in Solar Physics, 10, 5

\bibitem[{{Parker}(1983)}]{Parker1983}
{Parker}, E.~N. 1983, \apj, 264, 642

\bibitem[{{Parker}(1988)}]{Parker1988}
{Parker}, E.~N. 1988, \apj, 330, 474

\bibitem[{{Paschmann} {et~al.}(2005){Paschmann}, {Haaland}, {Sonnerup},
  {Hasegawa}, {Georgescu}, {Klecker}, {Phan}, {R{\`e}me}, \&
  {Vaivads}}]{Paschmann2005}
{Paschmann}, G., {Haaland}, S., {Sonnerup}, B.~U.~{\"O}., {et~al.} 2005,
  Annales Geophysicae, 23, 1481

\bibitem[{{Paschmann} \& {Sonnerup}(2008)}]{Paschmann2008}
{Paschmann}, G. \& {Sonnerup}, B.~U.~O. 2008, ISSI Scientific Reports Series,
  8, 65

\bibitem[{{Pecora} {et~al.}(2022){Pecora}, {Matthaeus}, {Primavera}, {Greco},
  {Chhiber}, {Bandyopadhyay}, \& {Servidio}}]{Pecora2022}
{Pecora}, F., {Matthaeus}, W.~H., {Primavera}, L., {et~al.} 2022, \apjl, 929,
  L10

\bibitem[{{Petschek}(1964)}]{Petschek1964}
{Petschek}, H.~E. 1964, in NASA Special Publication, Vol.~50, 425

\bibitem[{{Phan} {et~al.}(2020){Phan}, {Bale}, {Eastwood}, {Lavraud}, {Drake},
  {Oieroset}, {Shay}, {Pulupa}, {Stevens}, {MacDowall}, {Case}, {Larson},
  {Kasper}, {Whittlesey}, {Szabo}, {Korreck}, {Bonnell}, {de Wit}, {Goetz},
  {Harvey}, {Horbury}, {Livi}, {Malaspina}, {Paulson}, {Raouafi}, \&
  {Velli}}]{Phan2020}
{Phan}, T.~D., {Bale}, S.~D., {Eastwood}, J.~P., {et~al.} 2020, \apjs, 246, 34

\bibitem[{{Phan} {et~al.}(2021){Phan}, {Lavraud}, {Halekas}, {{\O}ieroset},
  {Drake}, {Eastwood}, {Shay}, {Pyakurel}, {Bale}, {Larson}, {Livi},
  {Whittlesey}, {Rahmati}, {Pulupa}, {McManus}, {Verniero}, {Bonnell},
  {Schwadron}, {Stevens}, {Case}, {Kasper}, {MacDowall}, {Szabo}, {Koval},
  {Korreck}, {Dudok de Wit}, {Malaspina}, {Goetz}, \& {Harvey}}]{Phan2021}
{Phan}, T.~D., {Lavraud}, B., {Halekas}, J.~S., {et~al.} 2021, \aap, 650, A13

\bibitem[{{Phan} {et~al.}(2013){Phan}, {Paschmann}, {Gosling}, {Oieroset},
  {Fujimoto}, {Drake}, \& {Angelopoulos}}]{Phan2013}
{Phan}, T.~D., {Paschmann}, G., {Gosling}, J.~T., {et~al.} 2013, \grl, 40, 11

\bibitem[{{Pontin}(2011)}]{Pontin2011}
{Pontin}, D.~I. 2011, Advances in Space Research, 47, 1508

\bibitem[{{Reisenfeld} {et~al.}(2001){Reisenfeld}, {Gary}, {Gosling},
  {Steinberg}, {McComas}, {Goldstein}, \& {Neugebauer}}]{Reisenfeld2001}
{Reisenfeld}, D.~B., {Gary}, S.~P., {Gosling}, J.~T., {et~al.} 2001, \jgr, 106,
  5693

\bibitem[{{Rosenbauer} {et~al.}(1977){Rosenbauer}, {Schwenn}, {Marsch},
  {Meyer}, {Miggenrieder}, {Montgomery}, {Muehlhaeuser}, {Pilipp}, {Voges}, \&
  {Zink}}]{Rosenbauer1977}
{Rosenbauer}, H., {Schwenn}, R., {Marsch}, E., {et~al.} 1977, Journal of
  Geophysics Zeitschrift Geophysik, 42, 561

\bibitem[{{Ruffolo} {et~al.}(2020){Ruffolo}, {Matthaeus}, {Chhiber}, {Usmanov},
  {Yang}, {Bandyopadhyay}, {Parashar}, {Goldstein}, {DeForest}, {Wan},
  {Chasapis}, {Maruca}, {Velli}, \& {Kasper}}]{Ruffolo2020}
{Ruffolo}, D., {Matthaeus}, W.~H., {Chhiber}, R., {et~al.} 2020, \apj, 902, 94

\bibitem[{{Schwadron} \& {McComas}(2021)}]{Schwardron2021}
{Schwadron}, N.~A. \& {McComas}, D.~J. 2021, \apj, 909, 95

\bibitem[{{Sonnerup} \& {Cahill}(1967)}]{Sonnerup1967}
{Sonnerup}, B.~U.~O. \& {Cahill}, L.~J., J. 1967, \jgr, 72, 171

\bibitem[{{Sonnerup} \& {Scheible}(1998)}]{Sonnerup1998}
{Sonnerup}, B.~U.~{\"O}. \& {Scheible}, M. 1998, ISSI Scientific Reports
  Series, 1, 185

\bibitem[{{Squire} {et~al.}(2020){Squire}, {Chandran}, \&
  {Meyrand}}]{Squire2020}
{Squire}, J., {Chandran}, B.~D.~G., \& {Meyrand}, R. 2020, 891, L2

\bibitem[{{Squire} {et~al.}(2022){Squire}, {Johnston}, {Mallet}, \&
  {Meyrand}}]{Squire2022}
{Squire}, J., {Johnston}, Z., {Mallet}, A., \& {Meyrand}, R. 2022, Physics of
  Plasmas, 29, 112903

\bibitem[{{Sterling} \& {Moore}(2020)}]{Sterling2020}
{Sterling}, A.~C. \& {Moore}, R.~L. 2020, \apjl, 896, L18

\bibitem[{{Teh} {et~al.}(2009){Teh}, {Sonnerup}, {Hu}, \& {Farrugia}}]{Teh2009}
{Teh}, W.~L., {Sonnerup}, B.~U.~{\"O}., {Hu}, Q., \& {Farrugia}, C.~J. 2009,
  Annales Geophysicae, 27, 807

\bibitem[{{Telloni} {et~al.}(2022){Telloni}, {Zank}, {Stangalini}, {Downs},
  {Liang}, {Nakanotani}, {Andretta}, {Antonucci}, {Sorriso-Valvo}, {Adhikari},
  {Zhao}, {Marino}, {Susino}, {Grimani}, {Fabi}, {D'Amicis}, {Perrone},
  {Bruno}, {Carbone}, {Mancuso}, {Romoli}, {Deppo}, {Fineschi}, {Heinzel},
  {Moses}, {Naletto}, {Nicolini}, {Spadaro}, {Teriaca}, {Frassati}, {Jerse},
  {Landini}, {Pancrazzi}, {Russano}, {Sasso}, {Biondo}, {Burtovoi}, {Capuano},
  {Casini}, {Casti}, {Chioetto}, {Leo}, {Giarrusso}, {Liberatore}, {Berghmans},
  {Auch{\`e}re}, {Cuadrado}, {Chitta}, {Harra}, {Kraaikamp}, {Long}, {Mandal},
  {Parenti}, {Pelouze}, {Peter}, {Rodriguez}, {Sch{\"u}hle}, {Schwanitz},
  {Smith}, {Verbeeck}, \& {Zhukov}}]{Telloni2022}
{Telloni}, D., {Zank}, G.~P., {Stangalini}, M., {et~al.} 2022, \apjl, 936, L25

\bibitem[{{Tenerani} {et~al.}(2021){Tenerani}, {Sioulas}, {Matteini},
  {Panasenco}, {Shi}, \& {Velli}}]{Tenerani2021}
{Tenerani}, A., {Sioulas}, N., {Matteini}, L., {et~al.} 2021, \apjl, 919, L31

\bibitem[{{Tenerani} {et~al.}(2020){Tenerani}, {Velli}, {Matteini},
  {R{\'e}ville}, {Shi}, {Bale}, {Kasper}, {Bonnell}, {Case}, {de Wit}, {Goetz},
  {Harvey}, {Klein}, {Korreck}, {Larson}, {Livi}, {MacDowall}, {Malaspina},
  {Pulupa}, {Stevens}, \& {Whittlesey}}]{Tenerani2020}
{Tenerani}, A., {Velli}, M., {Matteini}, L., {et~al.} 2020, \apjs, 246, 32

\bibitem[{{Zank} {et~al.}(2014){Zank}, {le Roux}, {Webb}, {Dosch}, \&
  {Khabarova}}]{Zank2014}
{Zank}, G.~P., {le Roux}, J.~A., {Webb}, G.~M., {Dosch}, A., \& {Khabarova}, O.
  2014, \apj, 797, 28

\end{thebibliography}

\end{document}